\documentclass[aps,twocolumn,showpacs,pra,groupedaddress,superscriptaddress,nofootinbib,longbibliography]{revtex4-1}
\usepackage{graphicx,amsmath,amssymb,amsbsy,subfigure,hyperref,bbm,times,txfonts}
\usepackage{color}
 \usepackage{graphicx}
 \usepackage{epstopdf}
\usepackage{amsmath}

\providecommand{\openone}{\leavevmode\hbox{\small1\kern-3.8pt\normalsize1}}

\makeatletter

\newcommand{\Rmnum}[1]{\expandafter\@slowromancap\romannumeral #1@}
\makeatother

\begin{document}

\title{Evidence of genuine quantum effects in nonequilibrium entropy production}

\author{Qing-Feng Xue}
\author{Xu-Cai Zhuang}
\author{De-Yang Duan}
\author{Ying-Jie Zhang}
\author{Wei-Bin Yan}
\author{Yun-Jie Xia}
\affiliation{School of Physics and Physical Engineering, Shandong Provincial Key Laboratory of Laser Polarization and Information Technology, Qufu Normal University, 273165, Qufu, China}
\author{Rosario Lo Franco}
\thanks{Email: rosario.lofranco@unipa.it}
\affiliation{Dipartimento di Ingegneria, Universit\`{a} degli Studi di Palermo,
Viale delle Scienze, 90128 Palermo, Italy}
\author{Zhong-Xiao Man}\thanks{Email: zxman@qfnu.edu.cn}
\affiliation{School of Physics and Physical Engineering, Shandong Provincial Key Laboratory of Laser Polarization and Information Technology, Qufu Normal University, 273165, Qufu, China}

\begin{abstract}
Entropy production is a fundamental concept that plays a crucial role in the second law of thermodynamics and the measure of irreversibility.
It imposes rigorous constraints on the kinds of transformations allowed in thermodynamic processes.
Using an optical setup, here we experimentally demonstrate the division of entropy production of an open quantum system into a population-related component and a coherence-related component, validating previous theoretical predictions.  
The coherence-related component represents a genuine quantum contribution with no classical counterpart.  
By adjusting bath temperatures and initial coherences of the system,
we first derive the total entropy production due to both populations and coherences, then remove all the coherences of the system to solely obtain the population-related contribution. The difference between these two results permits to isolate the coherence-related term. 
Based on this division, our experiment ultimately proves that irreversibility at the quantum level can be reduced through properly harnessing the two contributions to entropy production.  
\end{abstract}

\maketitle

\noindent The entropy changes of an open quantum system during a nonequilibrium process consist of two parts: one is the reversible entropy flow resulting from the system-environment
interaction, while the other one is the irreversible contribution, called entropy production \cite{RMP_Landi}.
Understanding and characterizing entropy production is a crucial task 
since it is closely related to the second law of thermodynamics,
and helps to characterize and measure the irreversibility of physical processes \cite{QT1,QT2,QT3,QT4,QT5,QT6}.
The irreversible entropy has practical implications for
assessing and designing the performance of thermal machines \cite{appli1,appli2}.
The discrepancy of an engine efficiency from Carnot's efficiency is directly related to entropy production, which suggests that the performance of a thermal machine may be improved by reducing the entropy production.

To advance theoretical understanding as well as technical applications, it is essential to identify the different contributions to entropy production in a nonequilibrium quantum process,
especially the involved genuine quantum effects  \cite{entpro1,entpro2,entpro3,entpro4,entpro5,entpro6}.
Recently, a remarkable achievement in quantum thermodynamics showed that the entropy production of an open quantum system,
subject to Markovian dynamics or thermal operations,
can be separated into terms depending on population imbalances and quantum coherences \cite{entpro1}.
Similarly, for a driven nonequilibrium quantum system, it was
pointed out that the irreversible work (and entropy)
can also be split into coherent and incoherent parts \cite{entpro4}.
The decomposition of entropy production unveils genuine quantum contributions and, furthermore, provides guidance for potential utilization of quantum coherence in thermodynamic tasks \cite{coh1,coh2,coh3,coh5,coh7,coh8,coh10,coh11,coh12,coh13,coh14,coh15,
coh17,coh18}.

Quantum coherence is a fundamental concept in quantum mechanics that also represents a key resource for quantum-enhanced technologies \cite{QI,l1}.
While it has long been anticipated that quantum coherence might have advantages in fields other than quantum information, especially in quantum thermodynamics, no general consensus has been reached so far \cite{Supercond}. 
There is an ongoing debate on this question, even regarding whether or not coherence can enhance the performances of quantum thermal machines \cite{Supercond,debate1,debate1,debate1,debate1,commoent1,commoent2,Nega1}.
In this context, the role of controlled coherences in increasing efficiency and power of an Otto heat engine has been investigated \cite{lubr}. 
In fact, different types of coherences have distinct effects on the entropy production. 
Vertical coherence, which is the coherence between different energy levels of a system, always leads to a positive contribution to entropy production under Markovian dynamics or thermal operations \cite{entpro1}.
In contrast, the coherence of degenerate levels, called horizontal coherence, can contribute negatively to entropy production because of the coupling of populations and coherences during the system dynamics \cite{commoent1,commoent2,Nega1}.
It has been demonstrated that horizontal coherence can enhance the energy flow without increasing irreversibility
\cite{Supercond}. Moreover, when the coherence is significant, the energy flow exhibits a scaling behavior, similar to a superconducting electric current \cite{Supercond}.

\begin{figure*}[t!]
	\begin{center}
		{\includegraphics[width=0.9\textwidth]{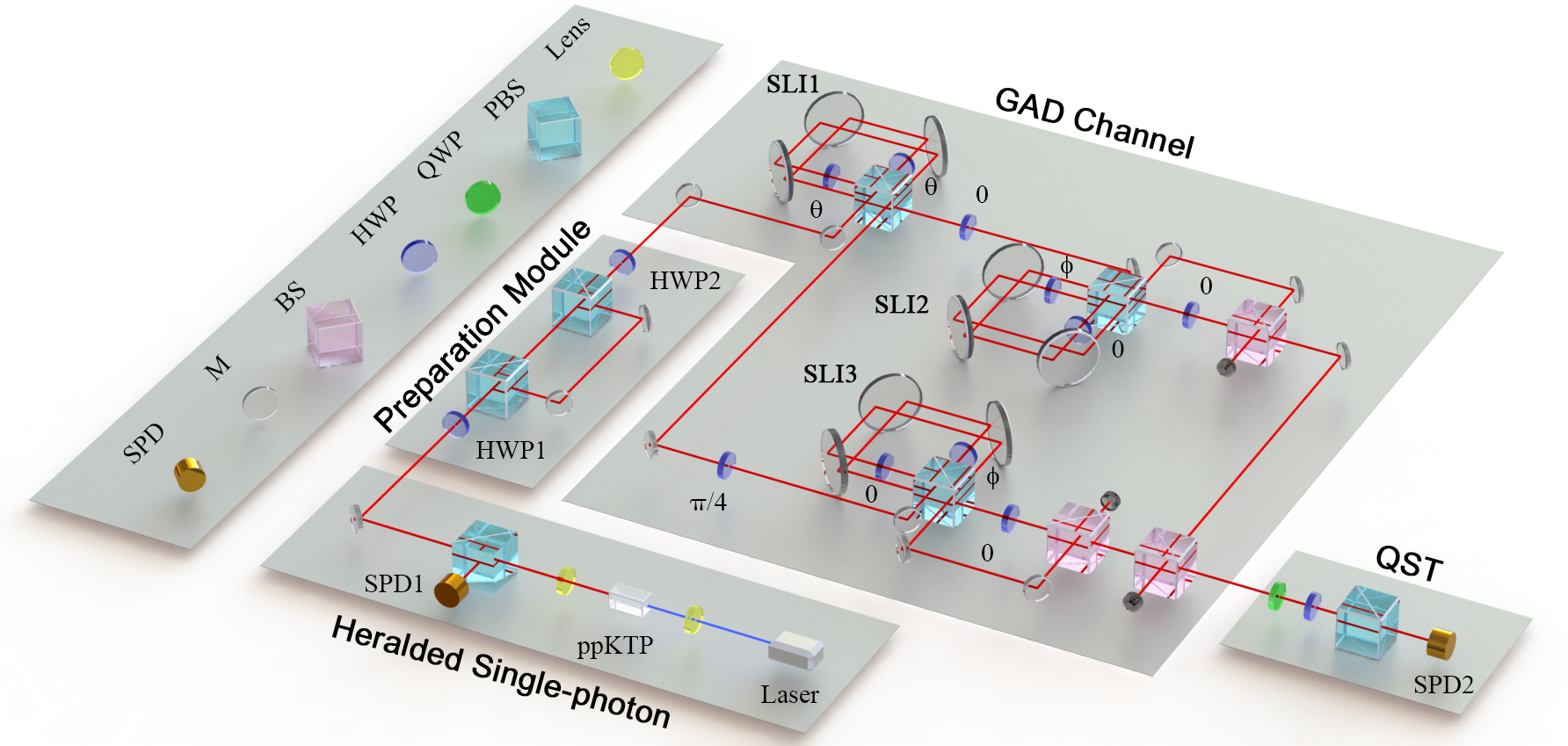} }
	\end{center}
	\caption{\textbf{Experimental setup.} 
		A continuous-wave laser at 405 nm generate photon pairs 
		from a ppKTP crystal, effectively working as a heralded single photon source when
		the trigger photons are detected by SPD1. The single photons firstly enter the
		preparation module which consists of two HWPs (i.e., HWP1 and HWP2), two mirrors, and two PBSs.
		The photons are prepared in the state $\rho_S$ of Eq.~(\ref{initstate}) and injected to the
		GAD channel which is implemented by three Sagnac-like interferometers (SLIs), named SLI1, SLI2 and SLI3,
		each one containing a PBS, two HWPs, and three mirrors. 
		After passing the GAD channel, the state $\rho_S$ of photons evolves into 
		$\rho_{S}^{\prime}$ (\ref{evol}). The last module is quantum state tomography (QST) for single photons,
		which includes a HWP, a QWP and a PBS.
		QWP: quarter wave plate, HWP: half wave plate, PBS: polarization beam splitter,
		BS: beam splitter, M: mirror, SPD: single photon detector.
	}	\label{model}
\end{figure*}

From the experimental perspective, entropy production has been so far observed in out-of-equilibrium intermediate scale quantum systems such as a Bose-Einstein condensate and a micromechanical resonator \cite{Brunelli_PRL}. Instead, the validation of genuine quantum contributions to entropy changes in elementary quantum systems has remained unexplored experimentally.
In this work, we employ a properly-assembled optical setup to verify in the lab the decomposition of entropy production into population-related and coherence-related components during the evolution of a two-level open quantum system. The findings confirm previous theoretical predictions \cite{entpro1}. Two scenarios are considered. In the first scenario, we measure the two components under various bath temperatures for a fixed initial state of the system. The dynamics of coherence is independent of the bath temperature, thus its contribution to the entropy production remains invariant. On the other hand, different bath temperatures lead to different equilibrium states of the system. Therefore, the population-related contribution varies with the bath temperature. In the second scenario, we prepare different initial states of the system with the same populations but different coherences, while keeping the bath temperature constant. In this case, the population-related component remains unchanged, while the coherence provides different contributions depending on the initial states.
We carry out two experiments in both scenarios: the first experiment reports the entropy production due to both populations and coherences; the second experiment determines the entropy production stemming solely from the populations of the system by removing all the coherences in the prepared initial state. The difference between these two results yields evidence of entropy production due to quantum coherence only.

\vskip0.6cm
\noindent\textbf{Results}\label{theory}

\noindent\textbf{Experimental system.}
We focus on a two-level system which undergoes a nonequlibrium quantum process due to the coupling with a thermal bath at temperature $T$.
The system is governed by the free Hamiltonian $\hat{H}_{S}
=\frac{\hbar\omega_{S}}{2}\hat{\sigma}_{z}$, with $\omega_{S}$ the transition frequency
and $\hat{\sigma}_{z}$ the Pauli $z$ operator.
In the stationary regime, 
the system reaches a thermal equilibrium state $\rho_{S}^{eq}=
\exp(-\beta\hat{H}_{S})/Z_{S}$, where $Z_{S}=\mathrm{tr}\left[\exp(-\beta\hat{H}_{S})\right]$
is the partition function and $\beta=1/k_{B}T$. We set $k_{B}=\hbar=1$ hereafter.

We use the horizontally (vertically) polarized state of photons, represented by $\left|H\right\rangle$ ($\left|V\right\rangle$),
to encode the ground (excited) state $\left|0\right\rangle$ ($\left|1\right\rangle$) of the open quantum system
of interest. The all-optical setup realized for the study is depicted in Fig.~\ref{model}. 
It essentially consists of four units: (i) the source of heralded single photons,
(ii) the preparation of initial states, (iii) the dynamics under generalized amplitude damping (GAD) channel, and (iv) the quantum state tomography. We recall that the GAD channel \cite{QI} represents the evolution of a two-level system (qubit) under a thermal bath (see Methods). 
In the preparation module, the signal photons pass through HWP1 with an angle $\alpha$,
which transforms the horizontal polarization state $\left| H \right\rangle$ 
into $\cos (2\alpha) \left| H \right\rangle + \sin (2\alpha) \left| V \right\rangle$.  
Photons are then divided into two paths after passing through the first PBS,
which overlap again via the combination of the second PBS
by precisely adjusting the fine-tuning knobs of the two mirror frames equipped with mirrors. 
The quantum coherence of the input state is destroyed when the path difference is set to be larger than the coherence length, and consequently the state of photons becomes $\cos^2(2\alpha)\left| H \right\rangle \left\langle H \right| + \sin^2(2\alpha)\left|V\right\rangle \left\langle V \right|$.
By setting the angle of HWP2 to be $\pi/8$, the state of photons is finally prepared to 
\begin{eqnarray}\label{initstate}
\rho_S&=&\frac{1}{2}\left( \left| H \right\rangle \left\langle H \right| + \left| V \right\rangle \left\langle V \right| \right) \nonumber\\
&&+ \frac{\cos (4\alpha)}{2} \left(\left| H \right\rangle \left\langle V \right| + \left| V \right\rangle \left\langle H \right|\right).
\end{eqnarray}
To measure the coherence of the photon state, we adopt the
$l_1$-norm of coherence $\widetilde{C}(\rho)$, which is defined as the sum of the absolute values
of all off-diagonal elements of a quantum state $\rho$, i.e.,
$\widetilde{C}(\rho) = \sum\limits_{i \ne j} \left| c _{ij} \right|$ for $\rho=\sum\limits_{i,j} c _{i,j} \left| i \right\rangle \left\langle j \right|$ \cite{l1}.
For the state $\rho_S$ of Eq.~(\ref{initstate}), we get $\widetilde{C}(\rho_{S})=\left|\cos(4\alpha)\right|$. 
Therefore, we can on-demand prepare states of photons with
various amounts of coherence by adjusting the angle $\alpha$ of HWP1.

After passing the GAD channel implemented by three
Sagnac-like interferometers (SLIs) (see Methods for
more details), the prepared photon state $\rho_S$ of Eq. (\ref{initstate}) evolves into 
\begin{equation}\label{evol}
\rho_{S}^{\prime}={p_g}\left| H \right\rangle \left\langle H \right|
+ {p_e}\left| V \right\rangle \left\langle V \right|
+ {p_c}\left(\left| H \right\rangle \left\langle V \right| + \left| V \right\rangle \left\langle H \right|\right), 
\end{equation}
with $p_g = pr + (1 - r)/2$, $p_e = (1 + r)/2-pr$,
and $p_c = \left[\cos (4\alpha) \sqrt {1-r}\right]/2 $. Notice that $p$ and $r$ are related, respectively, to the temperature $T$ of the bath and to the system-bath interaction time $t$. The range of $p\in[0.5,1]$ corresponds to $\beta\in[0,\infty]$, while $r\in[0,1]$ corresponds to $t\in[0,\infty]$. Our setup is designed such that the parameters $p$ and $r$ are adjustable by the angles of HWPs in the SLI1 and SLI2, SLI3, respectively (see Methods). 

The last module of the experimental setup performs single photon quantum tomography, which includes a HWP, a quarter-wave plate (QWP), and a PBS. We can completely reconstruct the single photon polarization state by performing projection measurements on four different projection bases ($\left| H \right\rangle$, $\left| V \right\rangle$, $\left| R \right\rangle$, $\left| D \right\rangle$), with $\left| R \right\rangle =(\left| H \right\rangle  + i\left| V \right\rangle )/{\sqrt 2 }$ and $\left| D \right\rangle = (\left| H \right\rangle  + \left| V \right\rangle )/{\sqrt 2 }$.

\noindent\textbf{Entropy production.}
In order to determine where entropy production comes from in a nonequlibrium quantum process,
one needs to analyze how the system evolves in time. The system dynamics usually involves two processes: changes of populations, that move the system closer to the equilibrium state,
and loss of quantum coherence when it is present in the beginning.
It is known that population imbalance is a classical source of entropy  \cite{class1,class2,class3,class4}, while it was only recently discovered that coherence represents a genuine quantum contribution to entropy production \cite{entpro1}.
In the following, we recall the decomposition of entropy production into two terms that are associated with population and coherence, respectively \cite{entpro1}. This theoretical prediction shall be validated by the proposed all-optical experiment.

For the dynamics in the GAD channel described in Eq.~(\ref{Kraus}) (reported in Methods),
one always has the contractive property $D(\rho_{S}^{\prime}\|\rho_{S}^{eq})\leq
D(\rho_{S}\|\rho_{S}^{eq})$, with $D(\rho_{1}\|\rho_{2})=\mathrm{tr}(\rho_{1}\ln\rho_{1}-\rho_{1}\ln\rho_{2})$ being
the relative entropy between states $\rho_{1}$ and $\rho_{2}$.
The entropy production can thus be defined based on the
relative entropy as 
\begin{equation}\label{EP}
\Sigma=D\left(\rho_{S}\|\rho_{S}^{eq}\right)-D\left(\rho_{S}^{\prime}\|\rho_{S}^{eq}\right)\geq0.
\end{equation}
It is known that the relative entropy can be split as
\begin{equation}\label{dec}
D\left(\rho_{S}\|\rho_{S}^{eq}\right)=D\left(\Delta (\rho_{S})\|\rho_{S}^{eq}\right)+C(\rho_{S}),
\end{equation}
where $\Delta (\rho_{S})$ denotes the dephasing map on $\rho_{S}$, 
removing all its coherences in the eigenbasis of $\hat{H}_{S}$,
while $C(\rho_{S})$ is the relative entropy of coherence given as
\begin{equation}\label{REC}
C(\rho_{S})=S\left(\Delta (\rho_{S})\right)-S(\rho_{S}),
\end{equation}
with $S(\rho)=-\mathrm{tr}\rho\ln\rho$ the von Neumann entropy of $\rho$.
By virtue of Eq.~(\ref{dec}), the entropy production $\Sigma$ defined in Eq.~(\ref{EP}) can be decomposed into two terms, one  due to the population and the other to the coherence, as \cite{entpro1}
\begin{equation}\label{EPdec}
\Sigma=\Sigma^{pop}+\Sigma^{coh},
\end{equation}
where
\begin{equation}\label{EPpop}
\Sigma^{pop}=D\left(\Delta (\rho_{S})\|\rho_{S}^{eq}\right)
-D\left(\Delta (\rho_{S}^{\prime})\|\rho_{S}^{eq}\right),
\end{equation}
and
\begin{equation}\label{EPcoh}
\Sigma^{coh}=C(\rho_{S})-C(\rho_{S}^{\prime}).
\end{equation}
The non-negativity of $\Sigma^{pop}$ is ensured by the fact that the evolution of population and coherence are independent of each other in the dynamical process being considered.
Also, $\Sigma^{coh}$ is always positive because coherence only decreases and is not generated in the considered evolution.
In the following, we experimentally demonstrate the 
formulations given in Eqs.~(\ref{EPdec})-(\ref{EPcoh}) and prove that the total entropy production can be reduced by decreasing only one of the two components $\Sigma^{pop}$ and $\Sigma^{coh}$.

\begin{figure*}[t!]
	\begin{center}
		{\includegraphics[width=0.32\linewidth]{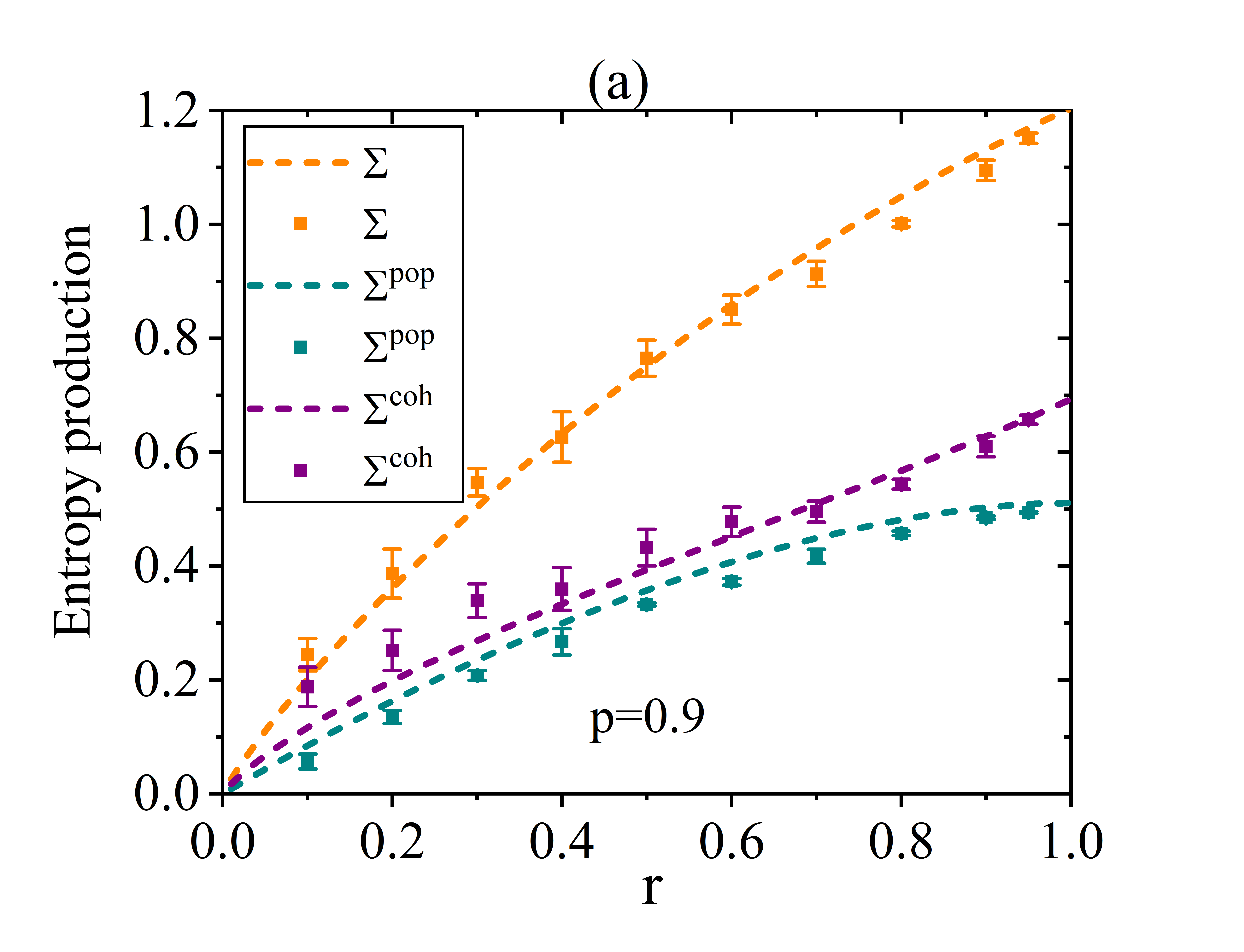} }
		{\includegraphics[width=0.32\linewidth]{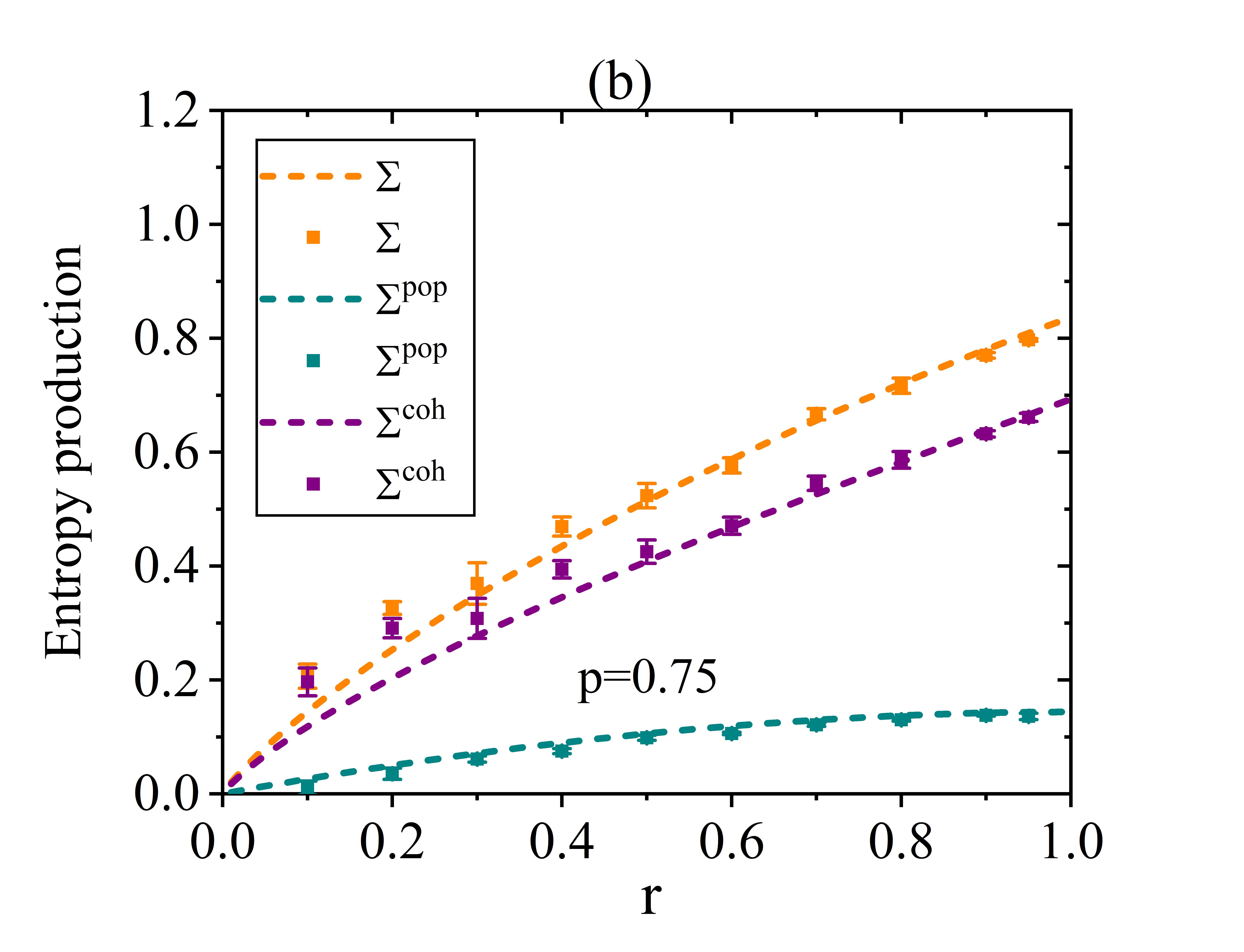} }
		{\includegraphics[width=0.32\linewidth]{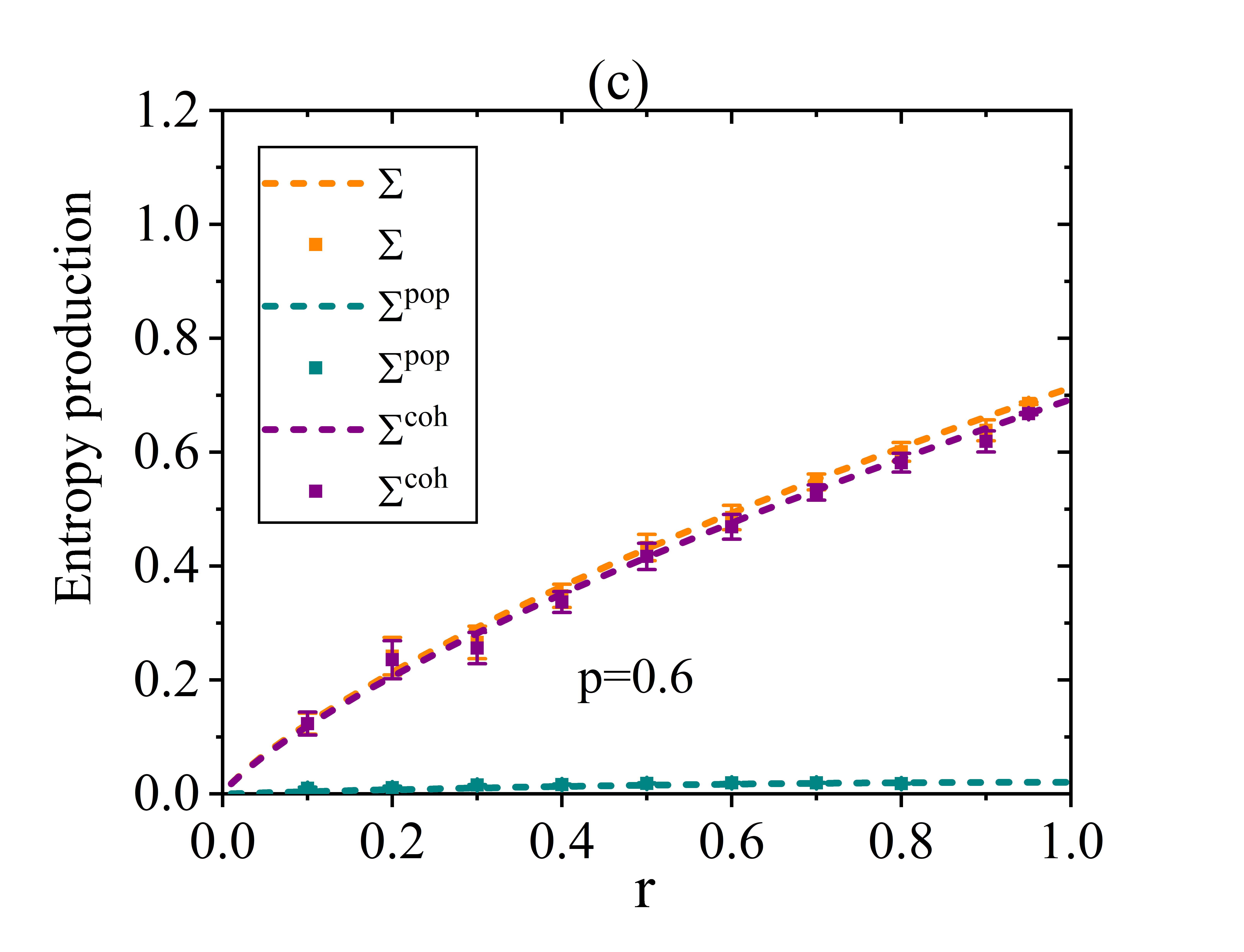} }
	\end{center}
	\caption{\textbf{Entropy production under different bath temperatures.} Total entropy production $\Sigma$ (orange curve and points) and its two components $\Sigma^{pop}$ (cyan curve and points) and $\Sigma^{coh}$ (purple curve and points) as a function of $r$ (system-bath interaction time) for different bath temperatures given in terms of $p$: (a) $p=0.9$, (b) $p=0.75$, (c) $p=0.6$.
		The initial state of photons is set to be $\rho_{S}=\left| D \right\rangle\left\langle D\right|$.
		The solid curves denote theoretical results and the dots represents the experimental data with their error bars. 
	}	\label{tem}
\end{figure*}

\noindent \textbf{Experimental observations.} We first demonstrate the decomposition of entropy production, given in Eqs.~(\ref{EPdec})-(\ref{EPcoh}),
for a fixed initial state of photons under different temperatures of the bath controlled by $p$. To determine the individual contributions of population and coherence to entropy production, we conduct two sets of experiments.   
Our first experiment deals with the scenario where the prepared system state has nonzero quantum coherence, from which we derive the total entropy production originating from both coherence and  population. Without loss of any generality, the photons are prepared in the state with maximum coherence $\widetilde{C}(\rho_S)=1$. To achieve this, we set the angle of HWP1 in the preparation module to be 0, which results in $\rho_{S}=\left| D \right\rangle\left\langle D\right|$ after the photons pass through the preparation module.
In the second experiment, we aim to obtain the entropy production contributed solely by the population
of state $\rho_{S}$. To this end, we need to remove all the coherences in the prepared state and transform 
$\rho_{S}$ into $\Delta(\rho_{S})=\frac{1}{2}\left(\left| H \right\rangle \left\langle H \right| + \left| V \right\rangle \left\langle V \right|\right)$. We can implement this process
by setting the angle of HWP1 in the preparation module to be $\pi/8$.
In each round of experiments, we choose three different temperatures of the bath given by, namely, $p=0.9$, $0.75$, and $0.6$. 
After reconstructing the evolved states in the two experimental scenarios via quantum tomography, we calculate the total entropy production $\Sigma$ and its population-related component $\Sigma^{pop}$. The contribution of quantum coherence $\Sigma^{coh}$ can be then determined by finding the difference between $\Sigma$ and $\Sigma^{pop}$.

\begin{figure*}[t!]
	\begin{center}
		{\includegraphics[width=0.32\linewidth]{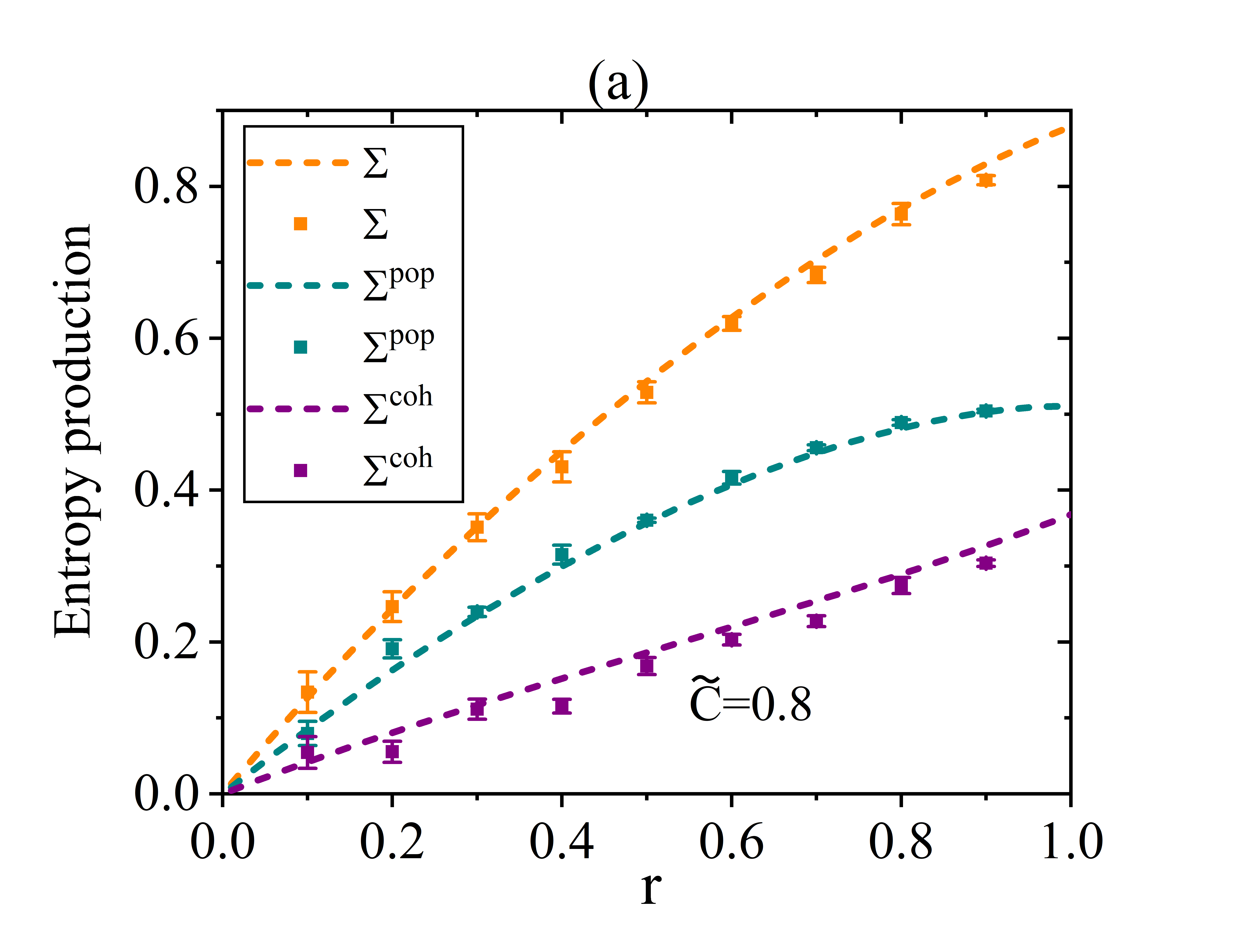} }
		{\includegraphics[width=0.32\linewidth]{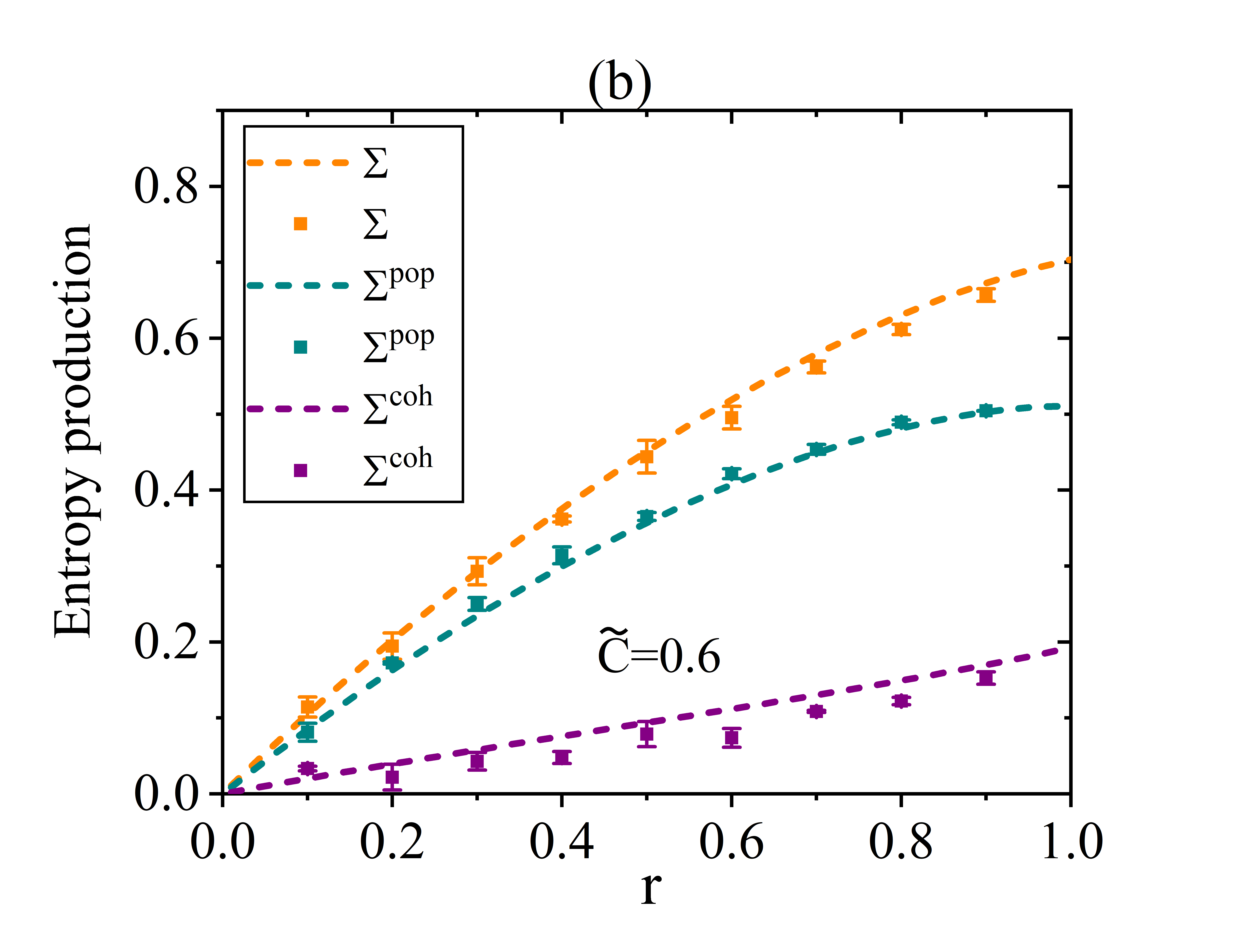} }
		{\includegraphics[width=0.32\linewidth]{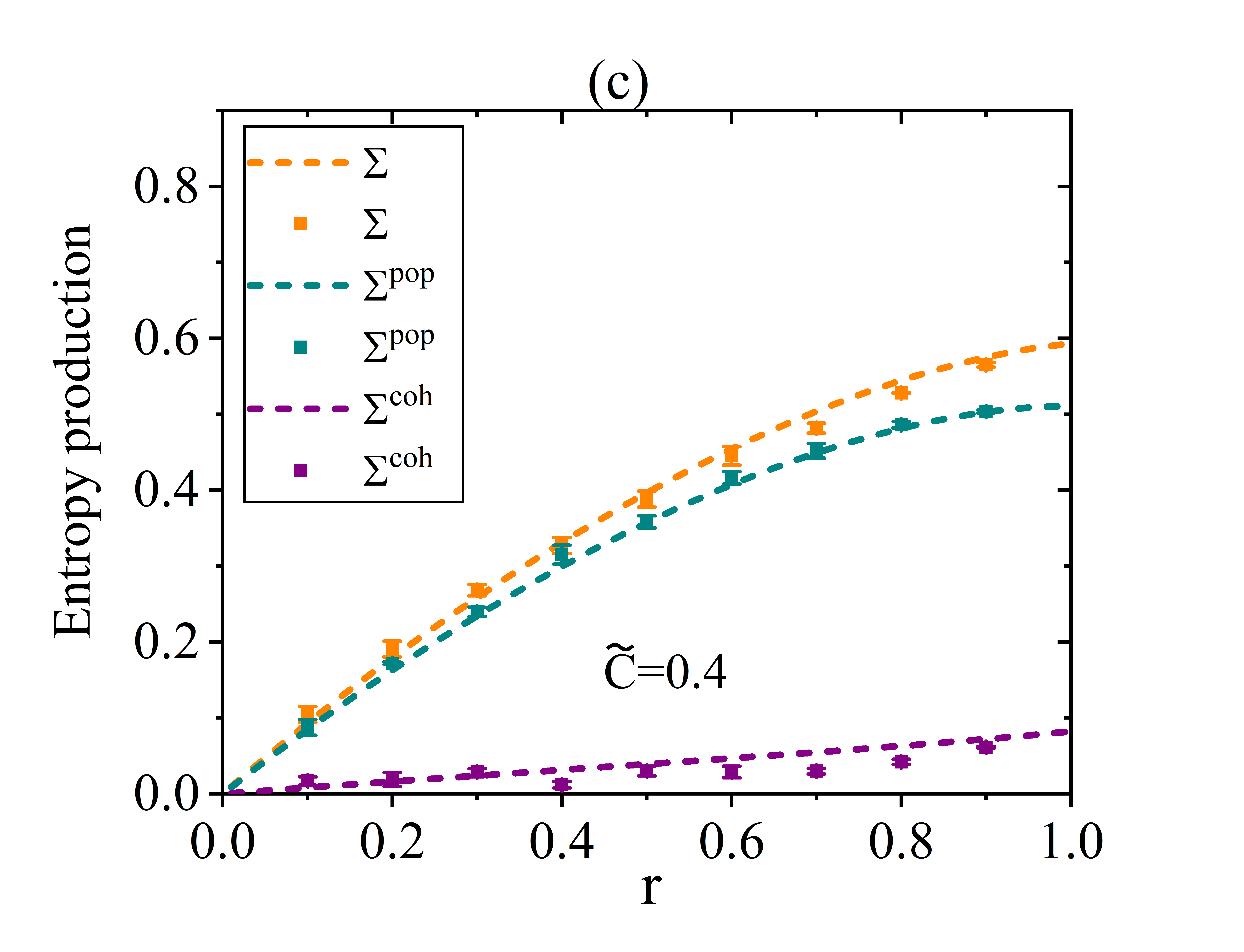} }
	\end{center}
	\caption{\textbf{Entropy production under different initial coherences.} Total entropy production $\Sigma$ (orange curve and points) and its two components
		$\Sigma^{pop}$ (cyan curve and points) and $\Sigma^{coh}$ (purple curve and points) as a function of $r$ (system-bath interaction time), with a fixed bath temperature $p=0.9$, for different amounts of initial coherence $\widetilde{C}$ in the prepared state.
		The solid curves denote theoretical results and the dots
		represents the experimental data with their error bars. 
	}	\label{init}
\end{figure*}

Our experimental results are compared with the theoretical predictions, as shown in Fig.~\ref{tem}, where panels (a), (b), and (c) correspond to $p=0.9$, $0.75$, and $0.6$, respectively.
We observe that the experimental data points are in agreement with the theoretical curves, which verifies that the entropy production of the system evolution comes from two sources: the population imbalances in a classical sense and the coherence embodying the genuine quantum effect. The role of population (coherence) in entropy production
can be highlighted by observing its behaviors under different bath temperature in terms of $p$.
The considered initial state after dephasing map reads
$\Delta(\rho_{S})=1/2\left(\left| H \right\rangle\left\langle H\right|+\left| V \right\rangle\left\langle V\right|\right)$,
which means that the system's temperatures is infinite due to the equal weights of the ground and excited states.
Therefore, as shown in Fig.~\ref{tem}(a), (b), and (c), when the temperature of the bath increases, that is $p$ gets smaller, 
the proportion of population contribution to entropy production decreases.
This result can be explained by the fact that the higher the bath temperature, the shorter the distance between the system initial state and its equilibrium state,
and consequently the smaller the entropy production induced by the dynamics of populations.
Specifically, when $p=0.6$, meaning that the bath temperature approaches infinite (or $\beta$ approaches zero), the population contribution to the entropy production almost vanishes,
while the coherence contribution nearly dominates the entire process of entropy production.
Since the dynamics of coherence is independent of the bath temperature, its contributions to entropy production in these three cases are identical, as evident in Fig.~\ref{tem}.

Next, we aim to prove the decomposition of entropy production by altering the coherence of the system initial states while keeping the system populations and the bath temperature constant. In other words, we intend to highlight the
contributions of distinct coherences to the entropy production under fixed contributions of populations.
This scenario complements the previous one, which was devoted a particular initial state with fixed coherence for different bath temperatures. In this way, we observe the phenomenon in a comprehensive view of possible system-bath conditions.  
Once again, we conduct two experiments as done above. 
In the first experiment, we obtain the total entropy production $\Sigma$ given by both populations and coherences.
We prepare three initial states $\rho _S$ of photons in the form of Eq.~(\ref{initstate}) 
with distinct coherences given as $\widetilde{C}({\rho_S})=0.8$, $\widetilde{C}({\rho_S})=0.6$, and $\widetilde{C}({\rho_S})=0.4$, respectively. This can be achieved by adjusting the angle $\alpha$ of HWP1 in the preparation module to $9.22^{\circ}$, $13.28^{\circ}$, and $16.61^{\circ}$, respectively. It is important to note that the populations of these three states are completely identical and have the same contributions to entropy production for a given bath temperature, e.g., $p=0.9$ used here.
In the second experiment, we remove the coherences of $\rho_S$ in the preparation stage 
and get the entropy production $\Sigma^{pop}$ stemming from populations only.
The entropy production $\Sigma^{coh}$ due to coherence is again obtained through the difference of  $\Sigma$ and $\Sigma^{pop}$. 

Panels (a), (b) and (c) of Fig.~\ref{init} display the experimental results for $\widetilde{C}({\rho_S})=0.8$, $\widetilde{C}({\rho_S})=0.6$, and $\widetilde{C}({\rho_S})=0.4$, respectively. These results are consistent with the theoretical predictions. 
We observe from panels (a), (b), and (c) of Fig.~\ref{init} that a smaller coherence in the system initial state leads to a smaller contribution $\Sigma^{coh}$ which, in turn, enables a reduction in the total entropy production $\Sigma$. In particular, as shown in Fig.~\ref{init}(c),
when the initial coherence is $\widetilde{C}=0.4$, which is a relatively small value, the contribution to the total entropy production $\Sigma$ is mainly supplied by the population-related component $\Sigma^{pop}$.

\vskip0.6cm
\noindent\textbf{Discussion}\label{concl}

\noindent In this work, we have experimentally demonstrated how entropy production in a nonequilibrium quantum process can be divided into population-related and coherence-related components. 
The contribution of coherence to entropy production is
a genuine quantum effect that only appears in the dynamics of open quantum systems.
By contrast, the population-related entropy production is completely equivalent to the classical case.
This discovery confirms that nonequilibrium processes in quantum systems are more complex than in classical thermodynamics. Hence, it is important to explore, from both theoretical and experimental perspective, the influence of quantum effects when studying quantum thermodynamics. 
For example, apart from coherence in the system of interest, future investigations include the influence of quantum resources present in nonequilibrium environments on the total entropy production \cite{Scully,NEE1,NEE2,NEE3,NEE4,NEE5,NEE6,NEE7,NEE8}.

Our experiments highlight the individual roles of both components in determining the total entropy production and also suggest possible ways to reduce it. Reducing entropy production equals to decrease the irreversibility of thermodynamic processes, a fundamental property with practical significance which can be used to enhance the performance of heat engines.
By adjusting the bath temperature, one can achieve an equilibrium state with a shorter distance to the system state, which will decrease the population-related entropy production and in turn the total entropy production. 
On the other hand, for a given bath temperature and system populations, by diminishing 
the system coherence, one can restrain the coherence-related entropy production and, at the same time, the total entropy production. Our research therefore indicates that the observation of different contributions to entropy production already has a value in potential applications.

We remark the effectiveness of the realized setup to provide a proof-of-principle of the phenomenon of entropy production at a quantum level, by suitably encoding system-bath interaction time and bath temperatures in optical elements of the apparatus. Quantum photonics and liner optics devices reveal a very suited platform to experimentally investigate quantum thermodynamic processes in principle. 

\vskip0.6cm
\noindent\textbf{Methods}

\noindent\textbf{Dynamics of the system.}
The evolution of an open quantum system in the presence of a thermal bath can be equivalently described by both the Kraus operator approach and the Markovian master equation.
The GAD channel involves both the relaxation and excitation processes and, for a two-level system (qubit), can be formulated by Kraus operators as \cite{QI}
\begin{equation}\label{Kraus}
\rho_{S}^{\prime}=\sum_{k=0}^{3} M_{k}\rho_{S}M_{k}^{\dag},
\end{equation}
in which
\begin{eqnarray}\label{KrausM}
&&M_{0}=\sqrt{p}\begin{pmatrix}
1  & 0\\
0         & \sqrt{1-r}
\end{pmatrix}, \hskip 0.5cm
M_{1}=\sqrt{p}\begin{pmatrix}
0  & \sqrt{r}\\
0         & 0
\end{pmatrix},\nonumber\\
&&M_{2}=\sqrt{1-p}\begin{pmatrix}
\sqrt{1-r}  & 0\\
0         & 1
\end{pmatrix},
M_{3}=\sqrt{1-p}\begin{pmatrix}
0  & 0\\
\sqrt{r}        & 0
\end{pmatrix},
\end{eqnarray}
where $M_{0}$ and $M_{1}$ ($M_{2}$ and $M_{3}$) correspond to the relaxation (excitation) processes, and $\rho_{S}$ ($\rho_{S}^{\prime}$)
representing the system state before (after) the interaction with the GAD channel.
In Eq.~(\ref{KrausM}), the parameters $p$ and $r$
characterize the GAD channel and their meanings
can be better understood by connecting the descriptions via
Kraus operators and master equation.
For a two-level system interacting with a thermal bath, the
master equation governing the system evolved state $\rho_{S}(t)$ reads \cite{breuer2002theory}
\begin{eqnarray}\label{ME}
\frac{d\rho_{S}(t)}{dt}&=&\gamma_{0}\left(\overline{n}+1\right)
\left[\hat{\sigma}_{-}\rho_{S}(t)\hat{\sigma}_{+}-\frac{1}{2}\{\hat{\sigma}_{+}\hat{\sigma}_{-},
\rho_{S}(t)\}\right]\nonumber\\
&&\gamma_{0}\overline{n}
\left[\hat{\sigma}_{+}\rho_{S}(t)\hat{\sigma}_{-}-
\frac{1}{2}\{\hat{\sigma}_{-}\hat{\sigma}_{+},\rho_{S}(t)\}\right],
\end{eqnarray}
where $\gamma_{0}$ is the rate of spontaneous emission,
$\hat{\sigma}_{+}=\left|1\right\rangle\left\langle0\right|$
($\hat{\sigma}_{-}=\left|0\right\rangle\left\langle1\right|$)
denotes the raising (lowering) operator
with $\left|1\right\rangle$ ($\left|0\right\rangle$) the excited (ground) state of the system, and $\overline{n}=[\exp(\omega_{S}/T)-1]^{-1}$ stands for the mean photon number at the frequency of the system.
These two descriptions of the system dynamics can be linked by means of the relations \cite{Kr}
\begin{equation}\label{r}
r=1-\exp\left[-(2\overline{n}+1)\gamma_{0}t\right],
\end{equation}
and
\begin{equation}\label{p}
p=\left[1+\exp\left(-\omega/T\right)\right]^{-1}.
\end{equation}
Namely, $r$ and $p$ are related, respectively, to the system-bath interaction time $t$ and to the temperature $T$ of the bath. Therefore, the range of $r\in[0,1]$ corresponds to that of $t\in[0,\infty]$, while $p\in[0.5,1]$ corresponds to $\beta\in[0,\infty]$.

\vskip0.6cm
\noindent\textbf{Experimental setup.} We employ a continuous-wave laser (with a central wavelength of 405 nm) to
generate photon pairs (with a central wavelength of 810 nm) from a periodically poled potassium titanium phosphate (ppKTP) nonlinear crystal by means of spontaneous parametric
down-conversion process. Photons with horizontal polarization are transmitted by the polarization beam splitter (PBS), whereas those with vertical polarization are reflected by PBS.
As a result, vertically polarized photons (i.e., the trigger) are detected by the single photon detector SPD1, while horizontally polarized photons (i.e., the signal) are directed to the main setup.

The GAD channel is realized by three Sagnac-like interferometers (SLIs), each one
containing a PBS, two HWPs, and three mirrors.
Through the first SLI (labeled as SLI1), by setting the angles of both HWPs
to $\theta$, we control the parameter $p = \cos^2 (2\theta)$ which is related to the bath temperature, as shown in Eq.~(\ref{p}).
Through the other two SLIs (named SLI2 and SLI3), by tuning one of the two HWPs to $\phi$ while keeping the other at $0$, the parameter $r=\sin^2 (2\phi)$ can be adjusted which is linked to the system-bath interaction time $t$, as shown in Eq.~(\ref{r}).
For example, to realize the excitation process represented by $M_2$ and $M_3$, the HWP
in the transmitted (reflected) path is set to be $\phi$ (0),
and the two outputs are combined by the beam splitter (BS) incoherently.
The relaxation process, represented by $M_0$ and $M_1$,
can be achieved by an inverse of the excitation process.
All paths are finally combined by the last BS incoherently.

\vskip0.6cm
\noindent\textbf{Data availability.} The data sets generated during and/or analyzed during the current study are available from the corresponding author on reasonable request.

\vskip0.6cm
\noindent\textbf{Acknowledgements}

\noindent This work was supported by National Natural Science Foundation (China) under Grant No. 11974209
and No. 12274257, Natural Science Foundation of Shandong Province (China) under Grant No. ZR2023LLZ015,
Taishan Scholar Project of Shandong Province (China) under Grant No. tsqn201812059,
and Youth Technological Innovation Support Program of Shandong Provincial Colleges and Universities under Grant
No. 2019KJJ015. 
R.L.F. acknowledges support from European Union -- NextGenerationEU -- grant MUR D.M. 737/2021 -- research project ``IRISQ''. 

\vskip0.6cm
\noindent\textbf{Author contributions}

\noindent Z.X.M. designed the experimental setup. Q.F.X., X.C.Z., D.Y.D. performed the experiment. R.L.F. supervised the theory behind the experiment. R.L.F. and Z.X.M. coordinated the project. All authors discussed the results and contributed to the preparation of the manuscript.

\vskip0.6cm
\noindent\textbf{Competing interests.} The authors declare no competing financial interests.


\begin{thebibliography}{53}%
\makeatletter
\providecommand \@ifxundefined [1]{%
 \@ifx{#1\undefined}
}%
\providecommand \@ifnum [1]{%
 \ifnum #1\expandafter \@firstoftwo
 \else \expandafter \@secondoftwo
 \fi
}%
\providecommand \@ifx [1]{%
 \ifx #1\expandafter \@firstoftwo
 \else \expandafter \@secondoftwo
 \fi
}%
\providecommand \natexlab [1]{#1}%
\providecommand \enquote  [1]{``#1''}%
\providecommand \bibnamefont  [1]{#1}%
\providecommand \bibfnamefont [1]{#1}%
\providecommand \citenamefont [1]{#1}%
\providecommand \href@noop [0]{\@secondoftwo}%
\providecommand \href [0]{\begingroup \@sanitize@url \@href}%
\providecommand \@href[1]{\@@startlink{#1}\@@href}%
\providecommand \@@href[1]{\endgroup#1\@@endlink}%
\providecommand \@sanitize@url [0]{\catcode `\\12\catcode `\$12\catcode
  `\&12\catcode `\#12\catcode `\^12\catcode `\_12\catcode `\%12\relax}%
\providecommand \@@startlink[1]{}%
\providecommand \@@endlink[0]{}%
\providecommand \url  [0]{\begingroup\@sanitize@url \@url }%
\providecommand \@url [1]{\endgroup\@href {#1}{\urlprefix }}%
\providecommand \urlprefix  [0]{URL }%
\providecommand \Eprint [0]{\href }%
\providecommand \doibase [0]{http://dx.doi.org/}%
\providecommand \selectlanguage [0]{\@gobble}%
\providecommand \bibinfo  [0]{\@secondoftwo}%
\providecommand \bibfield  [0]{\@secondoftwo}%
\providecommand \translation [1]{[#1]}%
\providecommand \BibitemOpen [0]{}%
\providecommand \bibitemStop [0]{}%
\providecommand \bibitemNoStop [0]{.\EOS\space}%
\providecommand \EOS [0]{\spacefactor3000\relax}%
\providecommand \BibitemShut  [1]{\csname bibitem#1\endcsname}%
\let\auto@bib@innerbib\@empty
\bibitem [{\citenamefont {Landi}\ and\ \citenamefont
  {Paternostro}(2021)}]{RMP_Landi}%
  \BibitemOpen
  \bibfield  {author} {\bibinfo {author} {\bibfnamefont {G.~T.}\ \bibnamefont
  {Landi}}\ and\ \bibinfo {author} {\bibfnamefont {M.}~\bibnamefont
  {Paternostro}},\ }\bibfield  {title} {\enquote {\bibinfo {title}
  {Irreversible entropy production: From classical to quantum},}\ }\href
  {\doibase 10.1103/RevModPhys.93.035008} {\bibfield  {journal} {\bibinfo
  {journal} {Rev. Mod. Phys.}\ }\textbf {\bibinfo {volume} {93}},\ \bibinfo
  {pages} {035008} (\bibinfo {year} {2021})}\BibitemShut {NoStop}%
\bibitem [{\citenamefont {Gemma}\ \emph {et~al.}(2004)\citenamefont {Gemma},
  \citenamefont {Michel},\ and\ \citenamefont {Mahler}}]{QT1}%
  \BibitemOpen
  \bibfield  {author} {\bibinfo {author} {\bibfnamefont {G.}~\bibnamefont
  {Gemma}}, \bibinfo {author} {\bibfnamefont {M.}~\bibnamefont {Michel}}, \
  and\ \bibinfo {author} {\bibfnamefont {G.}~\bibnamefont {Mahler}},\
  }\href@noop {} {\emph {\bibinfo {title} {Quantum Thermodynamics}}}\ (\bibinfo
   {publisher} {Berlin: Springer},\ \bibinfo {year} {2004})\BibitemShut
  {NoStop}%
\bibitem [{\citenamefont {Deffner}\ and\ \citenamefont {Campbell}(2019)}]{QT2}%
  \BibitemOpen
  \bibfield  {author} {\bibinfo {author} {\bibfnamefont {S.}~\bibnamefont
  {Deffner}}\ and\ \bibinfo {author} {\bibfnamefont {S.}~\bibnamefont
  {Campbell}},\ }\href {\doibase 10.1088/2053-2571/ab21c6} {\emph {\bibinfo
  {title} {Quantum Thermodynamics: An introduction to the thermodynamics of
  quantum information}}},\ 2053-2571\ (\bibinfo  {publisher} {Morgan and
  Claypool Publishers},\ \bibinfo {year} {2019})\BibitemShut {NoStop}%
\bibitem [{\citenamefont {Kosloff}(2013)}]{QT3}%
  \BibitemOpen
  \bibfield  {author} {\bibinfo {author} {\bibfnamefont {R.}~\bibnamefont
  {Kosloff}},\ }\bibfield  {title} {\enquote {\bibinfo {title} {Quantum
  thermodynamics: A dynamical viewpoint},}\ }\href {\doibase 10.3390/e15062100}
  {\bibfield  {journal} {\bibinfo  {journal} {Entropy}\ }\textbf {\bibinfo
  {volume} {15}},\ \bibinfo {pages} {2100} (\bibinfo {year}
  {2013})}\BibitemShut {NoStop}%
\bibitem [{\citenamefont {Vinjanampathy}\ and\ \citenamefont
  {Anders}(2016)}]{QT4}%
  \BibitemOpen
  \bibfield  {author} {\bibinfo {author} {\bibfnamefont {S.}~\bibnamefont
  {Vinjanampathy}}\ and\ \bibinfo {author} {\bibfnamefont {J.}~\bibnamefont
  {Anders}},\ }\bibfield  {title} {\enquote {\bibinfo {title} {Quantum
  thermodynamics},}\ }\href {\doibase 10.1080/00107514.2016.1201896} {\bibfield
   {journal} {\bibinfo  {journal} {Contemp. Phys.}\ }\textbf {\bibinfo {volume}
  {57}},\ \bibinfo {pages} {545} (\bibinfo {year} {2016})}\BibitemShut
  {NoStop}%
\bibitem [{\citenamefont {Goold}\ \emph {et~al.}(2016)\citenamefont {Goold},
  \citenamefont {Huber}, \citenamefont {Riera}, \citenamefont {del Rio},\ and\
  \citenamefont {Skrzypczyk}}]{QT5}%
  \BibitemOpen
  \bibfield  {author} {\bibinfo {author} {\bibfnamefont {J.}~\bibnamefont
  {Goold}}, \bibinfo {author} {\bibfnamefont {M.}~\bibnamefont {Huber}},
  \bibinfo {author} {\bibfnamefont {A.}~\bibnamefont {Riera}}, \bibinfo
  {author} {\bibfnamefont {L.}~\bibnamefont {del Rio}}, \ and\ \bibinfo
  {author} {\bibfnamefont {P.}~\bibnamefont {Skrzypczyk}},\ }\bibfield  {title}
  {\enquote {\bibinfo {title} {The role of quantum information in
  thermodynamics-a topical review},}\ }\href {\doibase
  10.1088/1751-8113/49/14/143001} {\bibfield  {journal} {\bibinfo  {journal}
  {J. Phys. A: Math. Theor.}\ }\textbf {\bibinfo {volume} {49}},\ \bibinfo
  {pages} {143001} (\bibinfo {year} {2016})}\BibitemShut {NoStop}%
\bibitem [{\citenamefont {Millen}\ and\ \citenamefont {Xuereb}(2016)}]{QT6}%
  \BibitemOpen
  \bibfield  {author} {\bibinfo {author} {\bibfnamefont {J.}~\bibnamefont
  {Millen}}\ and\ \bibinfo {author} {\bibfnamefont {A.}~\bibnamefont
  {Xuereb}},\ }\bibfield  {title} {\enquote {\bibinfo {title} {Perspective on
  quantum thermodynamics},}\ }\href {\doibase 10.1088/1367-2630/18/1/011002}
  {\bibfield  {journal} {\bibinfo  {journal} {New J. Phys.}\ }\textbf {\bibinfo
  {volume} {18}},\ \bibinfo {pages} {011002} (\bibinfo {year}
  {2016})}\BibitemShut {NoStop}%
\bibitem [{\citenamefont {Escher}\ \emph {et~al.}(2011)\citenamefont {Escher},
  \citenamefont {de~Matos~Filho},\ and\ \citenamefont {Davidovich}}]{appli1}%
  \BibitemOpen
  \bibfield  {author} {\bibinfo {author} {\bibfnamefont {B.~M.}\ \bibnamefont
  {Escher}}, \bibinfo {author} {\bibfnamefont {R.~L.}\ \bibnamefont
  {de~Matos~Filho}}, \ and\ \bibinfo {author} {\bibfnamefont {L.}~\bibnamefont
  {Davidovich}},\ }\bibfield  {title} {\enquote {\bibinfo {title} {General
  framework for estimating the ultimate precision limit in noisy
  quantum-enhanced metrology},}\ }\href {https://doi.org/10.1038/nphys1958}
  {\bibfield  {journal} {\bibinfo  {journal} {Nat. Phys.}\ }\textbf {\bibinfo
  {volume} {7}},\ \bibinfo {pages} {406} (\bibinfo {year} {2011})}\BibitemShut
  {NoStop}%
\bibitem [{\citenamefont {Verstraete}\ \emph {et~al.}(2009)\citenamefont
  {Verstraete}, \citenamefont {Wolf},\ and\ \citenamefont {Cirac}}]{appli2}%
  \BibitemOpen
  \bibfield  {author} {\bibinfo {author} {\bibfnamefont {F.}~\bibnamefont
  {Verstraete}}, \bibinfo {author} {\bibfnamefont {M.~M.}\ \bibnamefont
  {Wolf}}, \ and\ \bibinfo {author} {\bibfnamefont {J.~I.}\ \bibnamefont
  {Cirac}},\ }\bibfield  {title} {\enquote {\bibinfo {title} {Quantum
  computation and quantum-state engineering driven by dissipation},}\ }\href
  {\doibase https://doi.org/10.1038/nphys1342} {\bibfield  {journal} {\bibinfo
  {journal} {Nat. Phys.}\ }\textbf {\bibinfo {volume} {5}},\ \bibinfo {pages}
  {633} (\bibinfo {year} {2009})}\BibitemShut {NoStop}%
\bibitem [{\citenamefont {Santos}\ \emph {et~al.}(2019)\citenamefont {Santos},
  \citenamefont {C\'{e}leri}, \citenamefont {Landi},\ and\ \citenamefont
  {Paternostro}}]{entpro1}%
  \BibitemOpen
  \bibfield  {author} {\bibinfo {author} {\bibfnamefont {J.~P.}\ \bibnamefont
  {Santos}}, \bibinfo {author} {\bibfnamefont {L.~C.}\ \bibnamefont
  {C\'{e}leri}}, \bibinfo {author} {\bibfnamefont {G.~T.}\ \bibnamefont
  {Landi}}, \ and\ \bibinfo {author} {\bibfnamefont {M.}~\bibnamefont
  {Paternostro}},\ }\bibfield  {title} {\enquote {\bibinfo {title} {The role of
  quantum coherence in non-equilibrium entropy production},}\ }\href {\doibase
  10.1038/s41534-019-0138-y} {\bibfield  {journal} {\bibinfo  {journal} {npj
  Quantum Inf.}\ }\textbf {\bibinfo {volume} {5}},\ \bibinfo {pages} {23}
  (\bibinfo {year} {2019})}\BibitemShut {NoStop}%
\bibitem [{\citenamefont {Lostaglio}\ \emph {et~al.}(2015)\citenamefont
  {Lostaglio}, \citenamefont {Jennings},\ and\ \citenamefont
  {Rudolph}}]{entpro2}%
  \BibitemOpen
  \bibfield  {author} {\bibinfo {author} {\bibfnamefont {M.}~\bibnamefont
  {Lostaglio}}, \bibinfo {author} {\bibfnamefont {D.}~\bibnamefont {Jennings}},
  \ and\ \bibinfo {author} {\bibfnamefont {T.}~\bibnamefont {Rudolph}},\
  }\bibfield  {title} {\enquote {\bibinfo {title} {Description of quantum
  coherence in thermodynamic processes requires constraints beyond free
  energy},}\ }\href {https://doi.org/10.1038/ncomms7383} {\bibfield  {journal}
  {\bibinfo  {journal} {Nat. Commun.}\ }\textbf {\bibinfo {volume} {6}},\
  \bibinfo {pages} {6383} (\bibinfo {year} {2015})}\BibitemShut {NoStop}%
\bibitem [{\citenamefont {Mohammady}\ \emph {et~al.}(2020)\citenamefont
  {Mohammady}, \citenamefont {Aff\`{e}ves},\ and\ \citenamefont
  {Anders}}]{entpro3}%
  \BibitemOpen
  \bibfield  {author} {\bibinfo {author} {\bibfnamefont {M.~H.}\ \bibnamefont
  {Mohammady}}, \bibinfo {author} {\bibfnamefont {A.}~\bibnamefont
  {Aff\`{e}ves}}, \ and\ \bibinfo {author} {\bibfnamefont {J.}~\bibnamefont
  {Anders}},\ }\bibfield  {title} {\enquote {\bibinfo {title} {Energetic
  footprints of irreversibility in the quantum regime},}\ }\href
  {https://doi.org/10.1038/s42005-020-0356-9} {\bibfield  {journal} {\bibinfo
  {journal} {Commun. Phys.}\ }\textbf {\bibinfo {volume} {3}},\ \bibinfo
  {pages} {89} (\bibinfo {year} {2020})}\BibitemShut {NoStop}%
\bibitem [{\citenamefont {Francica}\ \emph {et~al.}(2019)\citenamefont
  {Francica}, \citenamefont {Goold},\ and\ \citenamefont {Plastina}}]{entpro4}%
  \BibitemOpen
  \bibfield  {author} {\bibinfo {author} {\bibfnamefont {G.}~\bibnamefont
  {Francica}}, \bibinfo {author} {\bibfnamefont {J.}~\bibnamefont {Goold}}, \
  and\ \bibinfo {author} {\bibfnamefont {F.}~\bibnamefont {Plastina}},\
  }\bibfield  {title} {\enquote {\bibinfo {title} {Role of coherence in the
  nonequilibrium thermodynamics of quantum systems},}\ }\href {\doibase
  10.1103/PhysRevE.99.042105} {\bibfield  {journal} {\bibinfo  {journal} {Phys.
  Rev. E}\ }\textbf {\bibinfo {volume} {99}},\ \bibinfo {pages} {042105}
  (\bibinfo {year} {2019})}\BibitemShut {NoStop}%
\bibitem [{\citenamefont {Varizi}\ \emph {et~al.}(2020)\citenamefont {Varizi},
  \citenamefont {Vieira}, \citenamefont {Cormick}, \citenamefont {Drumond},\
  and\ \citenamefont {Landi}}]{entpro5}%
  \BibitemOpen
  \bibfield  {author} {\bibinfo {author} {\bibfnamefont {A.~D.}\ \bibnamefont
  {Varizi}}, \bibinfo {author} {\bibfnamefont {A.~P.}\ \bibnamefont {Vieira}},
  \bibinfo {author} {\bibfnamefont {C.}~\bibnamefont {Cormick}}, \bibinfo
  {author} {\bibfnamefont {R.~C.}\ \bibnamefont {Drumond}}, \ and\ \bibinfo
  {author} {\bibfnamefont {G.~T.}\ \bibnamefont {Landi}},\ }\bibfield  {title}
  {\enquote {\bibinfo {title} {Quantum coherence and criticality in
  irreversible work},}\ }\href {\doibase 10.1103/PhysRevResearch.2.033279}
  {\bibfield  {journal} {\bibinfo  {journal} {Phys. Rev. Res.}\ }\textbf
  {\bibinfo {volume} {2}},\ \bibinfo {pages} {033279} (\bibinfo {year}
  {2020})}\BibitemShut {NoStop}%
\bibitem [{\citenamefont {Varizi}\ \emph {et~al.}(2021)\citenamefont {Varizi},
  \citenamefont {Cipolla}, \citenamefont {Perarnau-Llobet}, \citenamefont
  {Drumond},\ and\ \citenamefont {Landi}}]{entpro6}%
  \BibitemOpen
  \bibfield  {author} {\bibinfo {author} {\bibfnamefont {A.~D.}\ \bibnamefont
  {Varizi}}, \bibinfo {author} {\bibfnamefont {M.~A.}\ \bibnamefont {Cipolla}},
  \bibinfo {author} {\bibfnamefont {M.}~\bibnamefont {Perarnau-Llobet}},
  \bibinfo {author} {\bibfnamefont {R.~C.}\ \bibnamefont {Drumond}}, \ and\
  \bibinfo {author} {\bibfnamefont {G.~T.}\ \bibnamefont {Landi}},\ }\bibfield
  {title} {\enquote {\bibinfo {title} {Contributions from populations and
  coherences in non-equilibrium entropy production},}\ }\href {\doibase
  10.1088/1367-2630/abfe20} {\bibfield  {journal} {\bibinfo  {journal} {New J.
  Phys.}\ }\textbf {\bibinfo {volume} {23}},\ \bibinfo {pages} {063027}
  (\bibinfo {year} {2021})}\BibitemShut {NoStop}%
\bibitem [{\citenamefont {T\"{u}rkpen\c{c}e}\ and\ \citenamefont
  {M\"{u}stecapl{\i}o\v{g}lu}(2016)}]{coh1}%
  \BibitemOpen
  \bibfield  {author} {\bibinfo {author} {\bibfnamefont {D.}~\bibnamefont
  {T\"{u}rkpen\c{c}e}}\ and\ \bibinfo {author} {\bibfnamefont {\"{O}.~E.}\
  \bibnamefont {M\"{u}stecapl{\i}o\v{g}lu}},\ }\bibfield  {title} {\enquote
  {\bibinfo {title} {Quantum fuel with multilevel atomic coherence for
  ultrahigh specific work in a photonic carnot engine},}\ }\href {\doibase
  10.1103/PhysRevE.93.012145} {\bibfield  {journal} {\bibinfo  {journal} {Phys.
  Rev. E}\ }\textbf {\bibinfo {volume} {93}},\ \bibinfo {pages} {012145}
  (\bibinfo {year} {2016})}\BibitemShut {NoStop}%
\bibitem [{\citenamefont {Shi}\ \emph {et~al.}(2020)\citenamefont {Shi},
  \citenamefont {Shi}, \citenamefont {Wang}, \citenamefont {Hu}, \citenamefont
  {Liu}, \citenamefont {Yang},\ and\ \citenamefont {Fan}}]{coh2}%
  \BibitemOpen
  \bibfield  {author} {\bibinfo {author} {\bibfnamefont {Y.~H.}\ \bibnamefont
  {Shi}}, \bibinfo {author} {\bibfnamefont {H.~L.}\ \bibnamefont {Shi}},
  \bibinfo {author} {\bibfnamefont {X.~H.}\ \bibnamefont {Wang}}, \bibinfo
  {author} {\bibfnamefont {M.~L.}\ \bibnamefont {Hu}}, \bibinfo {author}
  {\bibfnamefont {S.~Y.}\ \bibnamefont {Liu}}, \bibinfo {author} {\bibfnamefont
  {W.~L.}\ \bibnamefont {Yang}}, \ and\ \bibinfo {author} {\bibfnamefont
  {H}~\bibnamefont {Fan}},\ }\bibfield  {title} {\enquote {\bibinfo {title}
  {Quantum coherence in a quantum heat engine},}\ }\href {\doibase
  10.1088/1751-8121/ab6a6b} {\bibfield  {journal} {\bibinfo  {journal} {J.
  Phys. A}\ }\textbf {\bibinfo {volume} {53}},\ \bibinfo {pages} {085301}
  (\bibinfo {year} {2020})}\BibitemShut {NoStop}%
\bibitem [{\citenamefont {Mehta}\ and\ \citenamefont {Johal}(2017)}]{coh3}%
  \BibitemOpen
  \bibfield  {author} {\bibinfo {author} {\bibfnamefont {V.}~\bibnamefont
  {Mehta}}\ and\ \bibinfo {author} {\bibfnamefont {R.~S.}\ \bibnamefont
  {Johal}},\ }\bibfield  {title} {\enquote {\bibinfo {title} {Quantum otto
  engine with exchange coupling in the presence of level degeneracy},}\ }\href
  {\doibase 10.1103/PhysRevE.96.032110} {\bibfield  {journal} {\bibinfo
  {journal} {Phys. Rev. E}\ }\textbf {\bibinfo {volume} {96}},\ \bibinfo
  {pages} {032110} (\bibinfo {year} {2017})}\BibitemShut {NoStop}%
\bibitem [{\citenamefont {Brandner}\ \emph {et~al.}(2015)\citenamefont
  {Brandner}, \citenamefont {Bauer}, \citenamefont {Schmid},\ and\
  \citenamefont {Seifert}}]{coh5}%
  \BibitemOpen
  \bibfield  {author} {\bibinfo {author} {\bibfnamefont {K.}~\bibnamefont
  {Brandner}}, \bibinfo {author} {\bibfnamefont {M.}~\bibnamefont {Bauer}},
  \bibinfo {author} {\bibfnamefont {M.~T}\ \bibnamefont {Schmid}}, \ and\
  \bibinfo {author} {\bibfnamefont {U.}~\bibnamefont {Seifert}},\ }\bibfield
  {title} {\enquote {\bibinfo {title} {Coherence-enhanced efficiency of
  feedback-driven quantum engines},}\ }\href {\doibase
  10.1088/1367-2630/17/6/065006} {\bibfield  {journal} {\bibinfo  {journal}
  {New J. Phys.}\ }\textbf {\bibinfo {volume} {17}},\ \bibinfo {pages} {065006}
  (\bibinfo {year} {2015})}\BibitemShut {NoStop}%
\bibitem [{\citenamefont {Niedenzu}\ \emph {et~al.}(2016)\citenamefont
  {Niedenzu}, \citenamefont {Gelbwaser-Klimovsky}, \citenamefont {Kofman},\
  and\ \citenamefont {Kurizki}}]{coh7}%
  \BibitemOpen
  \bibfield  {author} {\bibinfo {author} {\bibfnamefont {W.}~\bibnamefont
  {Niedenzu}}, \bibinfo {author} {\bibfnamefont {D.}~\bibnamefont
  {Gelbwaser-Klimovsky}}, \bibinfo {author} {\bibfnamefont {A.~G.}\
  \bibnamefont {Kofman}}, \ and\ \bibinfo {author} {\bibfnamefont
  {G.}~\bibnamefont {Kurizki}},\ }\bibfield  {title} {\enquote {\bibinfo
  {title} {On the operation of machines powered by quantum non-thermal
  baths},}\ }\href {\doibase 10.1088/1367-2630/18/8/083012} {\bibfield
  {journal} {\bibinfo  {journal} {New J. Phys.}\ }\textbf {\bibinfo {volume}
  {18}},\ \bibinfo {pages} {083012} (\bibinfo {year} {2016})}\BibitemShut
  {NoStop}%
\bibitem [{\citenamefont {Uzdin}(2016)}]{coh8}%
  \BibitemOpen
  \bibfield  {author} {\bibinfo {author} {\bibfnamefont {R.}~\bibnamefont
  {Uzdin}},\ }\bibfield  {title} {\enquote {\bibinfo {title} {Coherence-induced
  reversibility and collective operation of quantum heat machines via coherence
  recycling},}\ }\href {\doibase 10.1103/PhysRevApplied.6.024004} {\bibfield
  {journal} {\bibinfo  {journal} {Phys. Rev. Applied}\ }\textbf {\bibinfo
  {volume} {6}},\ \bibinfo {pages} {024004} (\bibinfo {year}
  {2016})}\BibitemShut {NoStop}%
\bibitem [{\citenamefont {Latune}\ \emph
  {et~al.}(2019{\natexlab{a}})\citenamefont {Latune}, \citenamefont
  {Sinayskiy},\ and\ \citenamefont {Petruccione}}]{coh10}%
  \BibitemOpen
  \bibfield  {author} {\bibinfo {author} {\bibfnamefont {C.L.}\ \bibnamefont
  {Latune}}, \bibinfo {author} {\bibfnamefont {I.}~\bibnamefont {Sinayskiy}}, \
  and\ \bibinfo {author} {\bibfnamefont {F.}~\bibnamefont {Petruccione}},\
  }\bibfield  {title} {\enquote {\bibinfo {title} {Quantum coherence, many-body
  correlations, and non-thermal effects for autonomous thermal machines},}\
  }\href {\doibase 10.1038/s41598-019-39300-4} {\bibfield  {journal} {\bibinfo
  {journal} {Sci. Rep.}\ }\textbf {\bibinfo {volume} {9}},\ \bibinfo {pages}
  {3191} (\bibinfo {year} {2019}{\natexlab{a}})}\BibitemShut {NoStop}%
\bibitem [{\citenamefont {{\AA}berg}(2014)}]{coh11}%
  \BibitemOpen
  \bibfield  {author} {\bibinfo {author} {\bibfnamefont {J.}~\bibnamefont
  {{\AA}berg}},\ }\bibfield  {title} {\enquote {\bibinfo {title} {Catalytic
  coherence},}\ }\href {\doibase 10.1103/PhysRevLett.113.150402} {\bibfield
  {journal} {\bibinfo  {journal} {Phys. Rev. Lett.}\ }\textbf {\bibinfo
  {volume} {113}},\ \bibinfo {pages} {150402} (\bibinfo {year}
  {2014})}\BibitemShut {NoStop}%
\bibitem [{\citenamefont {Korzekwa}\ \emph {et~al.}(2016)\citenamefont
  {Korzekwa}, \citenamefont {Lostaglio}, \citenamefont {Oppenheim},\ and\
  \citenamefont {Jennings}}]{coh12}%
  \BibitemOpen
  \bibfield  {author} {\bibinfo {author} {\bibfnamefont {K.}~\bibnamefont
  {Korzekwa}}, \bibinfo {author} {\bibfnamefont {M.}~\bibnamefont {Lostaglio}},
  \bibinfo {author} {\bibfnamefont {J.}~\bibnamefont {Oppenheim}}, \ and\
  \bibinfo {author} {\bibfnamefont {D.}~\bibnamefont {Jennings}},\ }\bibfield
  {title} {\enquote {\bibinfo {title} {The extraction of work from quantum
  coherence},}\ }\href {\doibase 10.1088/1367-2630/18/2/023045} {\bibfield
  {journal} {\bibinfo  {journal} {New J. Phys.}\ }\textbf {\bibinfo {volume}
  {18}},\ \bibinfo {pages} {023045} (\bibinfo {year} {2016})}\BibitemShut
  {NoStop}%
\bibitem [{\citenamefont {Li}\ \emph {et~al.}(2014)\citenamefont {Li},
  \citenamefont {Zou}, \citenamefont {Yu}, \citenamefont {Xu}, \citenamefont
  {Li},\ and\ \citenamefont {Shao}}]{coh13}%
  \BibitemOpen
  \bibfield  {author} {\bibinfo {author} {\bibfnamefont {H.}~\bibnamefont
  {Li}}, \bibinfo {author} {\bibfnamefont {J.}~\bibnamefont {Zou}}, \bibinfo
  {author} {\bibfnamefont {W.~L.}\ \bibnamefont {Yu}}, \bibinfo {author}
  {\bibfnamefont {B.~M.}\ \bibnamefont {Xu}}, \bibinfo {author} {\bibfnamefont
  {J.~G.}\ \bibnamefont {Li}}, \ and\ \bibinfo {author} {\bibfnamefont
  {B.}~\bibnamefont {Shao}},\ }\bibfield  {title} {\enquote {\bibinfo {title}
  {Quantum coherence rather than quantum correlations reflect the effects of a
  reservoir on a system's work capability},}\ }\href {\doibase
  10.1103/PhysRevE.89.052132} {\bibfield  {journal} {\bibinfo  {journal} {Phys.
  Rev. E}\ }\textbf {\bibinfo {volume} {89}},\ \bibinfo {pages} {052132}
  (\bibinfo {year} {2014})}\BibitemShut {NoStop}%
\bibitem [{\citenamefont {Liao}\ \emph {et~al.}(2010)\citenamefont {Liao},
  \citenamefont {Dong},\ and\ \citenamefont {Sun}}]{coh14}%
  \BibitemOpen
  \bibfield  {author} {\bibinfo {author} {\bibfnamefont {J.~Q.}\ \bibnamefont
  {Liao}}, \bibinfo {author} {\bibfnamefont {H.}~\bibnamefont {Dong}}, \ and\
  \bibinfo {author} {\bibfnamefont {C.~P.}\ \bibnamefont {Sun}},\ }\bibfield
  {title} {\enquote {\bibinfo {title} {Single-particle machine for quantum
  thermalization},}\ }\href {\doibase 10.1103/PhysRevA.81.052121} {\bibfield
  {journal} {\bibinfo  {journal} {Phys. Rev. A}\ }\textbf {\bibinfo {volume}
  {81}},\ \bibinfo {pages} {052121} (\bibinfo {year} {2010})}\BibitemShut
  {NoStop}%
\bibitem [{\citenamefont {Manatuly}\ \emph {et~al.}(2019)\citenamefont
  {Manatuly}, \citenamefont {Niedenzu}, \citenamefont {Rom\'an-Ancheyta},
  \citenamefont {\c{C}akmak}, \citenamefont {M\"ustecapl{\i}o\u{g}lu},\ and\
  \citenamefont {Kurizki}}]{coh15}%
  \BibitemOpen
  \bibfield  {author} {\bibinfo {author} {\bibfnamefont {A.}~\bibnamefont
  {Manatuly}}, \bibinfo {author} {\bibfnamefont {W.}~\bibnamefont {Niedenzu}},
  \bibinfo {author} {\bibfnamefont {R.}~\bibnamefont {Rom\'an-Ancheyta}},
  \bibinfo {author} {\bibfnamefont {B.}~\bibnamefont {\c{C}akmak}}, \bibinfo
  {author} {\bibfnamefont {\"O.~E.}\ \bibnamefont {M\"ustecapl{\i}o\u{g}lu}}, \
  and\ \bibinfo {author} {\bibfnamefont {G.}~\bibnamefont {Kurizki}},\
  }\bibfield  {title} {\enquote {\bibinfo {title} {Collectively enhanced
  thermalization via multiqubit collisions},}\ }\href {\doibase
  10.1103/PhysRevE.99.042145} {\bibfield  {journal} {\bibinfo  {journal} {Phys.
  Rev. E}\ }\textbf {\bibinfo {volume} {99}},\ \bibinfo {pages} {042145}
  (\bibinfo {year} {2019})}\BibitemShut {NoStop}%
\bibitem [{\citenamefont {Uzdin}\ \emph
  {et~al.}(2015{\natexlab{a}})\citenamefont {Uzdin}, \citenamefont {Levy},\
  and\ \citenamefont {Kosloff}}]{coh17}%
  \BibitemOpen
  \bibfield  {author} {\bibinfo {author} {\bibfnamefont {R.}~\bibnamefont
  {Uzdin}}, \bibinfo {author} {\bibfnamefont {A.}~\bibnamefont {Levy}}, \ and\
  \bibinfo {author} {\bibfnamefont {R.}~\bibnamefont {Kosloff}},\ }\bibfield
  {title} {\enquote {\bibinfo {title} {Equivalence of quantum heat machines,
  and quantum-thermodynamic signatures},}\ }\href {\doibase
  10.1103/PhysRevX.5.031044} {\bibfield  {journal} {\bibinfo  {journal} {Phys.
  Rev. X}\ }\textbf {\bibinfo {volume} {5}},\ \bibinfo {pages} {031044}
  (\bibinfo {year} {2015}{\natexlab{a}})}\BibitemShut {NoStop}%
\bibitem [{\citenamefont {Klatzow}\ \emph {et~al.}(2019)\citenamefont
  {Klatzow}, \citenamefont {Becker}, \citenamefont {Ledingham}, \citenamefont
  {Weinzetl}, \citenamefont {Kaczmarek}, \citenamefont {Saunders},
  \citenamefont {Nunn}, \citenamefont {Walmsley}, \citenamefont {Uzdin},\ and\
  \citenamefont {Poem}}]{coh18}%
  \BibitemOpen
  \bibfield  {author} {\bibinfo {author} {\bibfnamefont {J.}~\bibnamefont
  {Klatzow}}, \bibinfo {author} {\bibfnamefont {J.~N.}\ \bibnamefont {Becker}},
  \bibinfo {author} {\bibfnamefont {P.~M.}\ \bibnamefont {Ledingham}}, \bibinfo
  {author} {\bibfnamefont {C.}~\bibnamefont {Weinzetl}}, \bibinfo {author}
  {\bibfnamefont {K.~T.}\ \bibnamefont {Kaczmarek}}, \bibinfo {author}
  {\bibfnamefont {D.~J.}\ \bibnamefont {Saunders}}, \bibinfo {author}
  {\bibfnamefont {J.}~\bibnamefont {Nunn}}, \bibinfo {author} {\bibfnamefont
  {I.~A.}\ \bibnamefont {Walmsley}}, \bibinfo {author} {\bibfnamefont
  {R.}~\bibnamefont {Uzdin}}, \ and\ \bibinfo {author} {\bibfnamefont
  {E.}~\bibnamefont {Poem}},\ }\bibfield  {title} {\enquote {\bibinfo {title}
  {Experimental demonstration of quantum effects in the operation of
  microscopic heat engines},}\ }\href {\doibase 10.1103/PhysRevLett.122.110601}
  {\bibfield  {journal} {\bibinfo  {journal} {Phys. Rev. Lett.}\ }\textbf
  {\bibinfo {volume} {122}},\ \bibinfo {pages} {110601} (\bibinfo {year}
  {2019})}\BibitemShut {NoStop}%
\bibitem [{\citenamefont {Nielsen}\ and\ \citenamefont {Chuang}(2000)}]{QI}%
  \BibitemOpen
  \bibfield  {author} {\bibinfo {author} {\bibfnamefont {M.}~\bibnamefont
  {Nielsen}}\ and\ \bibinfo {author} {\bibfnamefont {I.}~\bibnamefont
  {Chuang}},\ }\href@noop {} {\emph {\bibinfo {title} {Quantum Computation and
  Quantum Information}}}\ (\bibinfo  {publisher} {Cambridge: Cambridge
  University Press},\ \bibinfo {year} {2000})\BibitemShut {NoStop}%
\bibitem [{\citenamefont {Streltsov}\ \emph {et~al.}(2017)\citenamefont
  {Streltsov}, \citenamefont {Adesso},\ and\ \citenamefont {Plenio}}]{l1}%
  \BibitemOpen
  \bibfield  {author} {\bibinfo {author} {\bibfnamefont {A.}~\bibnamefont
  {Streltsov}}, \bibinfo {author} {\bibfnamefont {G.}~\bibnamefont {Adesso}}, \
  and\ \bibinfo {author} {\bibfnamefont {M.~B.}\ \bibnamefont {Plenio}},\
  }\bibfield  {title} {\enquote {\bibinfo {title} {Colloquium: Quantum
  coherence as a resource},}\ }\href {\doibase 10.1103/RevModPhys.89.041003}
  {\bibfield  {journal} {\bibinfo  {journal} {Rev. Mod. Phys.}\ }\textbf
  {\bibinfo {volume} {89}},\ \bibinfo {pages} {041003} (\bibinfo {year}
  {2017})}\BibitemShut {NoStop}%
\bibitem [{\citenamefont {Tajima}\ and\ \citenamefont
  {Funo}(2021)}]{Supercond}%
  \BibitemOpen
  \bibfield  {author} {\bibinfo {author} {\bibfnamefont {H.}~\bibnamefont
  {Tajima}}\ and\ \bibinfo {author} {\bibfnamefont {K.}~\bibnamefont {Funo}},\
  }\bibfield  {title} {\enquote {\bibinfo {title} {Superconducting-like heat
  current: Effective cancellation of current-dissipation trade-off by quantum
  coherence},}\ }\href {\doibase 10.1103/PhysRevLett.127.190604} {\bibfield
  {journal} {\bibinfo  {journal} {Phys. Rev. Lett.}\ }\textbf {\bibinfo
  {volume} {127}},\ \bibinfo {pages} {190604} (\bibinfo {year}
  {2021})}\BibitemShut {NoStop}%
\bibitem [{\citenamefont {Uzdin}\ \emph
  {et~al.}(2015{\natexlab{b}})\citenamefont {Uzdin}, \citenamefont {Levy},\
  and\ \citenamefont {Kosloff}}]{debate1}%
  \BibitemOpen
  \bibfield  {author} {\bibinfo {author} {\bibfnamefont {R.}~\bibnamefont
  {Uzdin}}, \bibinfo {author} {\bibfnamefont {A.}~\bibnamefont {Levy}}, \ and\
  \bibinfo {author} {\bibfnamefont {R.}~\bibnamefont {Kosloff}},\ }\bibfield
  {title} {\enquote {\bibinfo {title} {Equivalence of quantum heat machines,
  and quantum-thermodynamic signatures},}\ }\href {\doibase
  10.1103/PhysRevX.5.031044} {\bibfield  {journal} {\bibinfo  {journal} {Phys.
  Rev. X}\ }\textbf {\bibinfo {volume} {5}},\ \bibinfo {pages} {031044}
  (\bibinfo {year} {2015}{\natexlab{b}})}\BibitemShut {NoStop}%
\bibitem [{\citenamefont {Latune}\ \emph
  {et~al.}(2019{\natexlab{b}})\citenamefont {Latune}, \citenamefont
  {Sinayskiy},\ and\ \citenamefont {Petruccione}}]{commoent1}%
  \BibitemOpen
  \bibfield  {author} {\bibinfo {author} {\bibfnamefont {C.~L.}\ \bibnamefont
  {Latune}}, \bibinfo {author} {\bibfnamefont {I.}~\bibnamefont {Sinayskiy}}, \
  and\ \bibinfo {author} {\bibfnamefont {F.}~\bibnamefont {Petruccione}},\
  }\bibfield  {title} {\enquote {\bibinfo {title} {Energetic and entropic
  effects of bath-induced coherences},}\ }\href {\doibase
  10.1103/PhysRevA.99.052105} {\bibfield  {journal} {\bibinfo  {journal} {Phys.
  Rev. A}\ }\textbf {\bibinfo {volume} {99}},\ \bibinfo {pages} {052105}
  (\bibinfo {year} {2019}{\natexlab{b}})}\BibitemShut {NoStop}%
\bibitem [{\citenamefont {Latune}\ \emph
  {et~al.}(2019{\natexlab{c}})\citenamefont {Latune}, \citenamefont
  {Sinayskiy},\ and\ \citenamefont {Petruccione}}]{commoent2}%
  \BibitemOpen
  \bibfield  {author} {\bibinfo {author} {\bibfnamefont {C.~L.}\ \bibnamefont
  {Latune}}, \bibinfo {author} {\bibfnamefont {I.}~\bibnamefont {Sinayskiy}}, \
  and\ \bibinfo {author} {\bibfnamefont {F.}~\bibnamefont {Petruccione}},\
  }\bibfield  {title} {\enquote {\bibinfo {title} {Thermodynamics from
  indistinguishability: Mitigating and amplifying the effects of the bath},}\
  }\href {\doibase 10.1103/PhysRevResearch.1.033192} {\bibfield  {journal}
  {\bibinfo  {journal} {Phys. Rev. Res.}\ }\textbf {\bibinfo {volume} {1}},\
  \bibinfo {pages} {033192} (\bibinfo {year} {2019}{\natexlab{c}})}\BibitemShut
  {NoStop}%
\bibitem [{\citenamefont {Latune}\ \emph {et~al.}(2020)\citenamefont {Latune},
  \citenamefont {Sinayskiy},\ and\ \citenamefont {Petruccione}}]{Nega1}%
  \BibitemOpen
  \bibfield  {author} {\bibinfo {author} {\bibfnamefont {C.~L.}\ \bibnamefont
  {Latune}}, \bibinfo {author} {\bibfnamefont {I.}~\bibnamefont {Sinayskiy}}, \
  and\ \bibinfo {author} {\bibfnamefont {F.}~\bibnamefont {Petruccione}},\
  }\bibfield  {title} {\enquote {\bibinfo {title} {Negative contributions to
  entropy production induced by quantum coherences},}\ }\href {\doibase
  10.1103/PhysRevA.102.042220} {\bibfield  {journal} {\bibinfo  {journal}
  {Phys. Rev. A}\ }\textbf {\bibinfo {volume} {102}},\ \bibinfo {pages}
  {042220} (\bibinfo {year} {2020})}\BibitemShut {NoStop}%
\bibitem [{\citenamefont {Camati}\ \emph {et~al.}(2019)\citenamefont {Camati},
  \citenamefont {Santos},\ and\ \citenamefont {Serra}}]{lubr}%
  \BibitemOpen
  \bibfield  {author} {\bibinfo {author} {\bibfnamefont {P.~A.}\ \bibnamefont
  {Camati}}, \bibinfo {author} {\bibfnamefont {Jonas F.~G.}\ \bibnamefont
  {Santos}}, \ and\ \bibinfo {author} {\bibfnamefont {R.~M.}\ \bibnamefont
  {Serra}},\ }\bibfield  {title} {\enquote {\bibinfo {title} {Coherence effects
  in the performance of the quantum otto heat engine},}\ }\href {\doibase
  10.1103/PhysRevA.99.062103} {\bibfield  {journal} {\bibinfo  {journal} {Phys.
  Rev. A}\ }\textbf {\bibinfo {volume} {99}},\ \bibinfo {pages} {062103}
  (\bibinfo {year} {2019})}\BibitemShut {NoStop}%
\bibitem [{\citenamefont {Brunelli}\ \emph {et~al.}(2018)\citenamefont
  {Brunelli}, \citenamefont {Fusco}, \citenamefont {Landig}, \citenamefont
  {Wieczorek}, \citenamefont {Hoelscher-Obermaier}, \citenamefont {Landi},
  \citenamefont {Semi\~ao}, \citenamefont {Ferraro}, \citenamefont {Kiesel},
  \citenamefont {Donner}, \citenamefont {De~Chiara},\ and\ \citenamefont
  {Paternostro}}]{Brunelli_PRL}%
  \BibitemOpen
  \bibfield  {author} {\bibinfo {author} {\bibfnamefont {M.}~\bibnamefont
  {Brunelli}}, \bibinfo {author} {\bibfnamefont {L.}~\bibnamefont {Fusco}},
  \bibinfo {author} {\bibfnamefont {R.}~\bibnamefont {Landig}}, \bibinfo
  {author} {\bibfnamefont {W.}~\bibnamefont {Wieczorek}}, \bibinfo {author}
  {\bibfnamefont {J.}~\bibnamefont {Hoelscher-Obermaier}}, \bibinfo {author}
  {\bibfnamefont {G.}~\bibnamefont {Landi}}, \bibinfo {author} {\bibfnamefont
  {F.~L.}\ \bibnamefont {Semi\~ao}}, \bibinfo {author} {\bibfnamefont
  {A.}~\bibnamefont {Ferraro}}, \bibinfo {author} {\bibfnamefont
  {N.}~\bibnamefont {Kiesel}}, \bibinfo {author} {\bibfnamefont
  {T.}~\bibnamefont {Donner}}, \bibinfo {author} {\bibfnamefont
  {G.}~\bibnamefont {De~Chiara}}, \ and\ \bibinfo {author} {\bibfnamefont
  {M.}~\bibnamefont {Paternostro}},\ }\bibfield  {title} {\enquote {\bibinfo
  {title} {Experimental determination of irreversible entropy production in
  out-of-equilibrium mesoscopic quantum systems},}\ }\href {\doibase
  10.1103/PhysRevLett.121.160604} {\bibfield  {journal} {\bibinfo  {journal}
  {Phys. Rev. Lett.}\ }\textbf {\bibinfo {volume} {121}},\ \bibinfo {pages}
  {160604} (\bibinfo {year} {2018})}\BibitemShut {NoStop}%
\bibitem [{\citenamefont {Tom\'e}\ and\ \citenamefont
  {de~Oliveira}(2010)}]{class1}%
  \BibitemOpen
  \bibfield  {author} {\bibinfo {author} {\bibfnamefont {T.}~\bibnamefont
  {Tom\'e}}\ and\ \bibinfo {author} {\bibfnamefont {M.~J.}\ \bibnamefont
  {de~Oliveira}},\ }\bibfield  {title} {\enquote {\bibinfo {title} {Entropy
  production in irreversible systems described by a fokker-planck equation},}\
  }\href {\doibase 10.1103/PhysRevE.82.021120} {\bibfield  {journal} {\bibinfo
  {journal} {Phys. Rev. E}\ }\textbf {\bibinfo {volume} {82}},\ \bibinfo
  {pages} {021120} (\bibinfo {year} {2010})}\BibitemShut {NoStop}%
\bibitem [{\citenamefont {Spinney}\ and\ \citenamefont {Ford}(2012)}]{class2}%
  \BibitemOpen
  \bibfield  {author} {\bibinfo {author} {\bibfnamefont {R.~E.}\ \bibnamefont
  {Spinney}}\ and\ \bibinfo {author} {\bibfnamefont {I.~J.}\ \bibnamefont
  {Ford}},\ }\bibfield  {title} {\enquote {\bibinfo {title} {Entropy production
  in full phase space for continuous stochastic dynamics},}\ }\href {\doibase
  10.1103/PhysRevE.85.051113} {\bibfield  {journal} {\bibinfo  {journal} {Phys.
  Rev. E}\ }\textbf {\bibinfo {volume} {85}},\ \bibinfo {pages} {051113}
  (\bibinfo {year} {2012})}\BibitemShut {NoStop}%
\bibitem [{\citenamefont {Schnakenberg}(1976)}]{class3}%
  \BibitemOpen
  \bibfield  {author} {\bibinfo {author} {\bibfnamefont {J.}~\bibnamefont
  {Schnakenberg}},\ }\bibfield  {title} {\enquote {\bibinfo {title} {Network
  theory of microscopic and macroscopic behavior of master equation systems},}\
  }\href {\doibase 10.1103/RevModPhys.48.571} {\bibfield  {journal} {\bibinfo
  {journal} {Rev. Mod. Phys.}\ }\textbf {\bibinfo {volume} {48}},\ \bibinfo
  {pages} {571--585} (\bibinfo {year} {1976})}\BibitemShut {NoStop}%
\bibitem [{\citenamefont {Tom\'e}\ and\ \citenamefont
  {de~Oliveira}(2012)}]{class4}%
  \BibitemOpen
  \bibfield  {author} {\bibinfo {author} {\bibfnamefont {T.}~\bibnamefont
  {Tom\'e}}\ and\ \bibinfo {author} {\bibfnamefont {M.~J.}\ \bibnamefont
  {de~Oliveira}},\ }\bibfield  {title} {\enquote {\bibinfo {title} {Entropy
  production in nonequilibrium systems at stationary states},}\ }\href
  {\doibase 10.1103/PhysRevLett.108.020601} {\bibfield  {journal} {\bibinfo
  {journal} {Phys. Rev. Lett.}\ }\textbf {\bibinfo {volume} {108}},\ \bibinfo
  {pages} {020601} (\bibinfo {year} {2012})}\BibitemShut {NoStop}%
\bibitem [{\citenamefont {Scully}\ \emph {et~al.}(2003)\citenamefont {Scully},
  \citenamefont {Zubairy}, \citenamefont {Agarwal},\ and\ \citenamefont
  {Walther}}]{Scully}%
  \BibitemOpen
  \bibfield  {author} {\bibinfo {author} {\bibfnamefont {M.~O.}\ \bibnamefont
  {Scully}}, \bibinfo {author} {\bibfnamefont {M.~S.}\ \bibnamefont {Zubairy}},
  \bibinfo {author} {\bibfnamefont {G.~S.}\ \bibnamefont {Agarwal}}, \ and\
  \bibinfo {author} {\bibfnamefont {H.}~\bibnamefont {Walther}},\ }\bibfield
  {title} {\enquote {\bibinfo {title} {Extracting work from a single heat bath
  via vanishing quantum coherence},}\ }\href {\doibase 10.1126/science.1078955}
  {\bibfield  {journal} {\bibinfo  {journal} {Science}\ }\textbf {\bibinfo
  {volume} {299}},\ \bibinfo {pages} {862} (\bibinfo {year}
  {2003})}\BibitemShut {NoStop}%
\bibitem [{\citenamefont {Leggio}\ \emph {et~al.}(2015)\citenamefont {Leggio},
  \citenamefont {Bellomo},\ and\ \citenamefont {Antezza}}]{NEE1}%
  \BibitemOpen
  \bibfield  {author} {\bibinfo {author} {\bibfnamefont {B.}~\bibnamefont
  {Leggio}}, \bibinfo {author} {\bibfnamefont {B.}~\bibnamefont {Bellomo}}, \
  and\ \bibinfo {author} {\bibfnamefont {M.}~\bibnamefont {Antezza}},\
  }\bibfield  {title} {\enquote {\bibinfo {title} {Quantum thermal machines
  with single nonequilibrium environments},}\ }\href {\doibase
  10.1103/PhysRevA.91.012117} {\bibfield  {journal} {\bibinfo  {journal} {Phys.
  Rev. A}\ }\textbf {\bibinfo {volume} {91}},\ \bibinfo {pages} {012117}
  (\bibinfo {year} {2015})}\BibitemShut {NoStop}%
\bibitem [{\citenamefont {Leggio}\ and\ \citenamefont {Antezza}(2016)}]{NEE2}%
  \BibitemOpen
  \bibfield  {author} {\bibinfo {author} {\bibfnamefont {B.}~\bibnamefont
  {Leggio}}\ and\ \bibinfo {author} {\bibfnamefont {M.}~\bibnamefont
  {Antezza}},\ }\bibfield  {title} {\enquote {\bibinfo {title} {Otto engine
  beyond its standard quantum limit},}\ }\href {\doibase
  10.1103/PhysRevE.93.022122} {\bibfield  {journal} {\bibinfo  {journal} {Phys.
  Rev. E}\ }\textbf {\bibinfo {volume} {93}},\ \bibinfo {pages} {022122}
  (\bibinfo {year} {2016})}\BibitemShut {NoStop}%
\bibitem [{\citenamefont {de~Assis}\ \emph {et~al.}(2019)\citenamefont
  {de~Assis}, \citenamefont {de~Mendon\c{c}a}, \citenamefont {Villas-Boas},
  \citenamefont {de~Souza}, \citenamefont {Sarthour}, \citenamefont
  {Oliveira},\ and\ \citenamefont {de~Almeida}}]{NEE3}%
  \BibitemOpen
  \bibfield  {author} {\bibinfo {author} {\bibfnamefont {R.~J.}\ \bibnamefont
  {de~Assis}}, \bibinfo {author} {\bibfnamefont {T.~M.}\ \bibnamefont
  {de~Mendon\c{c}a}}, \bibinfo {author} {\bibfnamefont {C.~J.}\ \bibnamefont
  {Villas-Boas}}, \bibinfo {author} {\bibfnamefont {A.~M.}\ \bibnamefont
  {de~Souza}}, \bibinfo {author} {\bibfnamefont {R.~S.}\ \bibnamefont
  {Sarthour}}, \bibinfo {author} {\bibfnamefont {I.~S.}\ \bibnamefont
  {Oliveira}}, \ and\ \bibinfo {author} {\bibfnamefont {N.~G.}\ \bibnamefont
  {de~Almeida}},\ }\bibfield  {title} {\enquote {\bibinfo {title} {Efficiency
  of a quantum otto heat engine operating under a reservoir at effective
  negative temperatures},}\ }\href {\doibase 10.1103/PhysRevLett.122.240602}
  {\bibfield  {journal} {\bibinfo  {journal} {Phys. Rev. Lett.}\ }\textbf
  {\bibinfo {volume} {122}},\ \bibinfo {pages} {240602} (\bibinfo {year}
  {2019})}\BibitemShut {NoStop}%
\bibitem [{\citenamefont {Cherubim}\ \emph {et~al.}(2019)\citenamefont
  {Cherubim}, \citenamefont {Brito},\ and\ \citenamefont {Deffner}}]{NEE4}%
  \BibitemOpen
  \bibfield  {author} {\bibinfo {author} {\bibfnamefont {C.}~\bibnamefont
  {Cherubim}}, \bibinfo {author} {\bibfnamefont {F.}~\bibnamefont {Brito}}, \
  and\ \bibinfo {author} {\bibfnamefont {S.}~\bibnamefont {Deffner}},\
  }\bibfield  {title} {\enquote {\bibinfo {title} {Non-thermal quantum engine
  in transmon qubits},}\ }\href {\doibase 10.3390/e21060545} {\bibfield
  {journal} {\bibinfo  {journal} {Entropy}\ }\textbf {\bibinfo {volume} {21}},\
  \bibinfo {pages} {545} (\bibinfo {year} {2019})}\BibitemShut {NoStop}%
\bibitem [{\citenamefont {Pezzutto}\ \emph {et~al.}(2019)\citenamefont
  {Pezzutto}, \citenamefont {Paternostro},\ and\ \citenamefont {Omar}}]{NEE5}%
  \BibitemOpen
  \bibfield  {author} {\bibinfo {author} {\bibfnamefont {M.}~\bibnamefont
  {Pezzutto}}, \bibinfo {author} {\bibfnamefont {M.}~\bibnamefont
  {Paternostro}}, \ and\ \bibinfo {author} {\bibfnamefont {Y.}~\bibnamefont
  {Omar}},\ }\bibfield  {title} {\enquote {\bibinfo {title} {An
  out-of-equilibrium non-{Markovian} quantum heat engine},}\ }\href {\doibase
  10.1088/2058-9565/aaf5b4} {\bibfield  {journal} {\bibinfo  {journal} {Quantum
  Sci. Technol.}\ }\textbf {\bibinfo {volume} {4}},\ \bibinfo {pages} {025002}
  (\bibinfo {year} {2019})}\BibitemShut {NoStop}%
\bibitem [{\citenamefont {Carollo}\ \emph {et~al.}(2020)\citenamefont
  {Carollo}, \citenamefont {Gambetta}, \citenamefont {Brandner}, \citenamefont
  {Garrahan},\ and\ \citenamefont {Lesanovsky}}]{NEE6}%
  \BibitemOpen
  \bibfield  {author} {\bibinfo {author} {\bibfnamefont {F.}~\bibnamefont
  {Carollo}}, \bibinfo {author} {\bibfnamefont {F.~M.}\ \bibnamefont
  {Gambetta}}, \bibinfo {author} {\bibfnamefont {K.}~\bibnamefont {Brandner}},
  \bibinfo {author} {\bibfnamefont {J.~P.}\ \bibnamefont {Garrahan}}, \ and\
  \bibinfo {author} {\bibfnamefont {I.}~\bibnamefont {Lesanovsky}},\ }\bibfield
   {title} {\enquote {\bibinfo {title} {Nonequilibrium quantum many-body
  {Rydberg} atom engine},}\ }\href {\doibase 10.1103/PhysRevLett.124.170602}
  {\bibfield  {journal} {\bibinfo  {journal} {Phys. Rev. Lett.}\ }\textbf
  {\bibinfo {volume} {124}},\ \bibinfo {pages} {170602} (\bibinfo {year}
  {2020})}\BibitemShut {NoStop}%
\bibitem [{\citenamefont {Rom{\'{a}}n-Ancheyta}\ \emph
  {et~al.}(2019)\citenamefont {Rom{\'{a}}n-Ancheyta}, \citenamefont
  {\c{C}akmak},\ and\ \citenamefont {M\"{u}stecapl{\i}o{\u{g}}lu}}]{NEE7}%
  \BibitemOpen
  \bibfield  {author} {\bibinfo {author} {\bibfnamefont {R.}~\bibnamefont
  {Rom{\'{a}}n-Ancheyta}}, \bibinfo {author} {\bibfnamefont {B.}~\bibnamefont
  {\c{C}akmak}}, \ and\ \bibinfo {author} {\bibfnamefont {\"{O}.~E.}\
  \bibnamefont {M\"{u}stecapl{\i}o{\u{g}}lu}},\ }\bibfield  {title} {\enquote
  {\bibinfo {title} {Spectral signatures of non-thermal baths in quantum
  thermalization},}\ }\href {\doibase 10.1088/2058-9565/ab5e4f} {\bibfield
  {journal} {\bibinfo  {journal} {Quantum Sci. Technol.}\ }\textbf {\bibinfo
  {volume} {5}},\ \bibinfo {pages} {015003} (\bibinfo {year}
  {2019})}\BibitemShut {NoStop}%
\bibitem [{\citenamefont {Niedenzu}\ \emph {et~al.}(2018)\citenamefont
  {Niedenzu}, \citenamefont {Mukherjee}, \citenamefont {Ghosh}, \citenamefont
  {Kofman},\ and\ \citenamefont {Kurizki}}]{NEE8}%
  \BibitemOpen
  \bibfield  {author} {\bibinfo {author} {\bibfnamefont {W.}~\bibnamefont
  {Niedenzu}}, \bibinfo {author} {\bibfnamefont {V.}~\bibnamefont {Mukherjee}},
  \bibinfo {author} {\bibfnamefont {A.}~\bibnamefont {Ghosh}}, \bibinfo
  {author} {\bibfnamefont {A.~G.}\ \bibnamefont {Kofman}}, \ and\ \bibinfo
  {author} {\bibfnamefont {G.}~\bibnamefont {Kurizki}},\ }\bibfield  {title}
  {\enquote {\bibinfo {title} {Quantum engine efficiency bound beyond the
  second law of thermodynamics},}\ }\href {\doibase 10.1038/s41467-017-01991-6}
  {\bibfield  {journal} {\bibinfo  {journal} {Nat. Commu.}\ }\textbf {\bibinfo
  {volume} {9}},\ \bibinfo {pages} {165} (\bibinfo {year} {2018})}\BibitemShut
  {NoStop}%
\bibitem [{\citenamefont {Breuer}\ and\ \citenamefont
  {Petruccione}(2002)}]{breuer2002theory}%
  \BibitemOpen
  \bibfield  {author} {\bibinfo {author} {\bibfnamefont {H.~P.}\ \bibnamefont
  {Breuer}}\ and\ \bibinfo {author} {\bibfnamefont {F.}~\bibnamefont
  {Petruccione}},\ }\href@noop {} {\emph {\bibinfo {title} {The theory of open
  quantum systems}}}\ (\bibinfo  {publisher} {Oxford: Oxford University
  Press},\ \bibinfo {year} {2002})\BibitemShut {NoStop}%
\bibitem [{\citenamefont {Zou}\ \emph {et~al.}(2017)\citenamefont {Zou},
  \citenamefont {Li}, \citenamefont {Wang}, \citenamefont {Cao}, \citenamefont
  {Ren}, \citenamefont {Yin}, \citenamefont {Peng}, \citenamefont {Wang},\ and\
  \citenamefont {Pan}}]{Kr}%
  \BibitemOpen
  \bibfield  {author} {\bibinfo {author} {\bibfnamefont {W.-J.}\ \bibnamefont
  {Zou}}, \bibinfo {author} {\bibfnamefont {Y.-H.}\ \bibnamefont {Li}},
  \bibinfo {author} {\bibfnamefont {S.-C.}\ \bibnamefont {Wang}}, \bibinfo
  {author} {\bibfnamefont {Y.}~\bibnamefont {Cao}}, \bibinfo {author}
  {\bibfnamefont {J.-G.}\ \bibnamefont {Ren}}, \bibinfo {author} {\bibfnamefont
  {J.}~\bibnamefont {Yin}}, \bibinfo {author} {\bibfnamefont {C.-Z.}\
  \bibnamefont {Peng}}, \bibinfo {author} {\bibfnamefont {X.-B.}\ \bibnamefont
  {Wang}}, \ and\ \bibinfo {author} {\bibfnamefont {J.-W.}\ \bibnamefont
  {Pan}},\ }\bibfield  {title} {\enquote {\bibinfo {title} {Protecting
  entanglement from finite-temperature thermal noise via weak measurement and
  quantum measurement reversal},}\ }\href {\doibase 10.1103/PhysRevA.95.042342}
  {\bibfield  {journal} {\bibinfo  {journal} {Phys. Rev. A}\ }\textbf {\bibinfo
  {volume} {95}},\ \bibinfo {pages} {042342} (\bibinfo {year}
  {2017})}\BibitemShut {NoStop}%
\end{thebibliography}

%

\end{document}